\documentclass[acmsmall]{acmart}

\pdfoutput=1

\usepackage{graphicx} 
\usepackage{booktabs}
\usepackage{multirow}
\usepackage{listings}
\usepackage{xcolor}
\usepackage{subcaption}

\usepackage{hyperref}

\usepackage{tcolorbox}
\definecolor{findgray}{HTML}{EFEFEF} 
\definecolor{findline}{HTML}{888888} 

\newtcolorbox{findingbox}{
    colback=findgray,      
    colframe=findline,     
    boxrule=0pt,           
    leftrule=4pt,          
    sharp corners,         
    arc=0pt,               
    left=6pt, right=6pt, top=6pt, bottom=6pt, 
    parbox=false           
}
\usepackage{tikz}

\usetikzlibrary{shapes.multipart}  
\usetikzlibrary{shadows}           
\usetikzlibrary{arrows.meta}       
\usetikzlibrary{positioning}       
\usetikzlibrary{backgrounds}       
\usetikzlibrary{fit}
\usepackage{pgfplots}
\pgfplotsset{compat=1.18}

\newcommand{\phead}[1]{\vspace{1mm}\noindent\textbf{#1}~~~}

\title{An Empirical Study of Speculative Decoding on Software Engineering Tasks}

\author{Yijia Li}
\affiliation{%
  \institution{Zhejiang University}
  \country{China}}
\email{yijia.li@zju.edu.cn}

\author{Junkai Chen}
\affiliation{%
  \institution{Singapore Management University}
  \country{Singapore}}
\email{junkaichen@smu.edu.sg}

\author{Xing Hu}
\authornote{Corresponding author.}
\affiliation{%
  \institution{Zhejiang University}
  \country{China}}
\email{xinghu@zju.edu.cn}

\author{Xin Xia}
\affiliation{%
  \institution{Zhejiang University}
  \country{China}}

\affiliation{%
  \institution{Hangzhou High-Tech Zone (Binjiang) Institute of Blockchain and Data Security}
  \country{China}}
\email{xin.xia@acm.org}

\date{\today}

\setcopyright{cc}
\setcctype{by}
\acmJournal{PACMSE}
\acmYear{2026} \acmVolume{3} \acmNumber{ISSTA} \acmArticle{ISSTA054}
\acmMonth{10} \acmDOI{10.1145/3832145}

\begin{document}

\begin{abstract}
Large Language Models (LLMs) have become widely used for Software Engineering (SE) tasks, spanning from function-level code generation to complex repository-level workflows.
However, the high latency of autoregressive inference remains a significant bottleneck, hindering their deployment in interactive environments.
While Speculative Decoding (SD) offers a promising technique for lossless acceleration, prior research on long-context repository-level tasks and complex agentic interactions remains limited.
To bridge this gap, we present a systematic empirical study to evaluate the effectiveness of SD in SE tasks.
We benchmark a representative spectrum of strategies, encompassing both model-based and model-free methods, across generation, editing, and repair scenarios.
Our empirical results show that SD provides clear acceleration potential for SE tasks, but its realized benefits jointly vary with model architecture and task scenario.
Specifically, model-based approaches are well-suited for code generation, whereas model-free methods are better adapted to repository-level repair and editing scenarios.
We further observe that the repetitiveness of SE tasks improves the performance of model-free methods, while complex agentic workflows can introduce repetitive failure modes that skew acceleration measurements.
In contrast to natural language tasks, the higher predictability of SE tasks allows for more aggressive hyperparameter settings.
Our findings provide practical guidance for selecting, configuring, and evaluating SD methods in SE scenarios.
\end{abstract}



\maketitle

\section{Introduction}

The capabilities of Large Language Models (LLMs) have advanced significantly during the past few years, leading to a paradigm shift across various domains, including Software Engineering (SE)~\cite{hou2024large}. 
In the early stages of LLM development around 2023, their application in SE was typically limited to relatively simple tasks, such as function-level code generation~\cite{chen2021evaluatinglargelanguagemodels, austin2021programsynthesislargelanguage}; 
by contrast, we observed that state-of-the-art LLMs in late 2025 could even handle long-horizon SE tasks, like solving complex GitHub issues~\cite{jimenez2024swebench,deng2025swe}. 
Assisted by techniques including LLM agents~\cite{wang2024openhands,yang2024swe} and reinforcement learning~\cite{guo2025deepseek,jaech2024openai}, these models are enabled to think, observe, and reflect by inference-time scaling (i.e., generating more output tokens) to improve performance. 
For instance, Qwen3-32B improves its LiveCodeBench Pass@1 from 31.3\% to 65.7\% by generating extended reasoning traces to verify and refine its logic~\cite{qwen3technicalreport}.

As LLMs generate more tokens to solve tasks, response latency has become a major bottleneck for user experience.
Therefore, many researchers have been exploring how to improve the speed of this generation process (i.e., inference acceleration of LLMs)~\cite{khoshnoodi2024comprehensivesurveyacceleratedgeneration} to mitigate this issue. 
For example, parameter quantization~\cite{frantar2022gptq,dettmers2022gpt3,lin2024awq} and model pruning~\cite{frantar2023sparsegpt} accelerate LLM inference by simplifying the architecture and weights of models to reduce the computation requirements, thereby accelerating their generation. 
Among these solutions, Speculative Decoding (SD)~\cite{leviathan2023fast,chen2023accelerating} has emerged as a promising technique by optimizing the decoding process of LLMs. 
Specifically, it operates on a ``\textit{draft-verify}'' paradigm, where a lightweight proxy of the original model first rapidly generates multiple draft tokens as candidates, which are then verified in parallel by the target model. The technique of SD presents many advantages over other techniques, e.g., it is easy to integrate without modification of model weights and does not degrade the generation quality by design. 
Therefore, it is widely adopted and researched~\cite{zheng2023judgingllmasajudgemtbenchchatbot,cobbe2021trainingverifierssolvemath,alpaca,hermann2015teachingmachinesreadcomprehend,xia2024unlockingefficiencylargelanguage} in the field of natural language. 
In the context of SE, although there are some works exploring the adoption and improvement of SD in SE~\cite{zhaofastcoder,peidingefficientedit}, it remains unknown how mainstream SD methods perform on popular SE tasks and LLMs.

To bridge the gap, this paper presents the first systematic empirical evaluation of different SD approaches on SE tasks. 
Specifically, we design the following research questions (RQs) to guide our study:
\begin{itemize}
    \item \textbf{RQ1: How effective is SD on various SE tasks?}

    In this research question (RQ), we would like to uncover the overall effectiveness of SD approaches on SE tasks. Specifically, we select four representative SD methods (e.g., Eagle-3~\cite{li2025eagle3scalinginferenceacceleration}) and experiment with them on diverse SE benchmarks, including six benchmarks spanning code generation, repair, and editing. These approaches and benchmarks are widely adopted by various entities and reasonably representative.

    \item \textbf{RQ2: How do model scale and architecture impact acceleration?}

    In this RQ, we investigate how model scale and architecture affect the effectiveness of SD. By evaluating both dense and sparsely activated MoE models spanning diverse scales, we analyze how model scale, activation patterns, and generation quality influence SD performance. Through this exploration, we aim to provide insights on how to better evaluate SD in SE.  

    \item \textbf{RQ3: How does the performance of SD differ between natural language and code?}

    This RQ presents a comparative analysis between SE and natural language tasks on their inference acceleration. 
    We try to explore the performance difference of SD and understand why this discrepancy exists from the perspective of language modeling and the acceleration principle.
\end{itemize}

We obtain the following findings when answering the RQs:
\begin{itemize}

    \item SD demonstrates clear acceleration potential across SE tasks, but its realized benefits are affected by task category and target-model characteristics, including scale and architecture.

    \item We find that the widespread ``infinite loops'' in complex agentic tasks introduce significant noise when evaluating the performance of SD.

    \item The repetitiveness of SE tasks improves the performance of model-free methods. In contrast to general natural language tasks, the higher predictability of SE tasks allows for more aggressive hyperparameters (i.e., a larger number of draft tokens).
\end{itemize}

Our contributions are summarized as follows:
\begin{itemize}
    \item We present a large-scale empirical study of SD on SE tasks. To the best of our knowledge, it is the first systematic empirical study of SD in the SE community.
    
    \item We conduct an in-depth analysis of the acceleration performance under various settings, e.g., different LLMs, SD methods, tasks, and parameters. Valuable takeaways are provided based on our analyses and discussion. 
    
    \item We provide insights and implications to practitioners in both academia and industry. These findings shed light on future deployment and research directions on SD for SE.
\end{itemize}

\phead{Paper Organization.} 
Section~\ref{sec:background} introduces the background of SD and related works on LLM and SE. 
Section~\ref{sec:methodology} discusses our research methodology, including studied methods, benchmarks, etc.
Section~\ref{sec:evaluation} describes the results of our experiments. 
Section~\ref{sec:discussion} discusses the impact of hyperparameters, implications, and threats to validity of our paper.
Section~\ref{sec:conclusion} concludes the paper.

\section{Background and Related Work}
\label{sec:background}

The application of LLMs in SE has been widely investigated through various empirical studies. Existing literature predominantly focuses on evaluating model capabilities across the software development lifecycle, ranging from replicating empirical research workflows~\cite{liang2024can} to code generation and automated program repair~\cite{huang2023empirical,sun2025empirical,shang2025large}. Researchers have also analyzed how execution information and prompt engineering enhance the functional correctness of generated code~\cite{jiang2023impact,zhou2024out}. However, these prior evaluations mainly concentrate on functional correctness and generation quality, using metrics such as repair rates and test pass rates. As LLMs are increasingly integrated into interactive development environments, inference latency becomes a critical bottleneck that directly affects user experience, motivating a closer examination of inference efficiency in SE scenarios.

To address the deployment requirements of LLMs, various inference acceleration techniques have been proposed, including parameter quantization, model pruning, and knowledge distillation. However, these techniques often introduce trade-offs in model performance and cannot guarantee lossless acceleration. Recent empirical studies have started to investigate such acceleration techniques in SE~\cite{wang2025empirical,afrin2025quantization,melin2024precision,haque2025quantization,siam2026exploring}. In contrast, SD has emerged as a promising lossless acceleration strategy that maintains the original output distribution. Yet, previous evaluations of SD have primarily been conducted on general natural language benchmarks, leaving its efficiency and behavioral characteristics on complex SE tasks largely unexplored. Our study bridges this gap by providing a systematic empirical evaluation of SD in SE scenarios.

To provide the necessary context for our study, this section first introduces the representative paradigms of SD, and then reviews commonly used SE benchmarks that motivate our benchmark selection.

\subsection{Speculative Decoding}
\label{subsec:sd_formulation}

SD accelerates LLM inference by employing a ``\textit{draft-verify}'' paradigm~\cite{chen2023accelerating, leviathan2023fast}. Instead of generating tokens one by one, it first uses a lightweight method to tentatively predict a sequence of future tokens (the ``\textit{drafting}'' phase). The large target model whose inference we aim to accelerate then validates these predictions in parallel (the ``\textit{verifying}'' phase). This strategy works because standard inference requires generating tokens serially. In contrast, the verification phase allows the target model to process an entire sequence of drafted tokens in parallel, with a latency roughly equivalent to generating a single token.
As an effective optimization for LLM inference, SD is widely applicable to autoregressive models and has demonstrated widespread utility, being integrated into mainstream inference engines such as vLLM~\cite{kwon2023efficient} and NVIDIA TensorRT-LLM~\cite{nvidia2024tensorrtllm}.
Note that in this paper, we only discuss the SD technique on autoregressive, Transformer-based language models.

Formally, the SD framework involves a target model $M_q$ parameterized by $\theta_q$ and a specialized efficient draft model $M_p$. In this process, given a context sequence $x_{<t}$, the draft model $M_p$ first efficiently generates a candidate chain $\hat{x}_{t:t+\gamma}$ of length $\gamma$, which is subsequently validated by the verifier $M_q$ in a single forward pass via a verification mechanism to ensure the output is lossless, in other words, mathematically identical to what the target model would generate alone.

In the greedy decoding setting, the verification simplifies to a strict equality check. A candidate token $\hat{x}_{t+i}$ at step $i$ (where $0 \le i < \gamma$) is accepted if and only if it aligns exactly with the top-1 prediction of the target model $M_q$:
\begin{equation}
\label{eq:greedy}  
\hat{x}_{t+i} = \operatorname{argmax}_{x} P_{M_q}(x \mid x_{<t}, \hat{x}_{t:t+i-1}).
\end{equation}
Upon the first mismatch at index $i$ (i.e., rejection),  the rejected draft token $\hat{x}_{t+i}$ is discarded and replaced by the correct token $x_{t+i}$, which is derived directly from the target model's logits computed during the parallel verification phase. All subsequent candidate tokens are also discarded.

For stochastic settings (i.e., Temperature $> 0$), SD employs speculative sampling~\cite{leviathan2023fast, chen2023accelerating} to guarantee that the generation remains lossless. Given that the experiments in this study are conducted under the greedy decoding setting, we omit a detailed discussion of this technique.

Based on the techniques used for drafting, we categorize existing approaches into two primary categories: \textit{Model-Based} and \textit{Model-Free}.

\subsubsection{Model-Based Methods}
This category utilizes a smaller neural model to approximate the target distribution of $M_q$.
Standard SD methods~\cite{chen2023accelerating, leviathan2023fast} employ a separate small language model to generate drafts for a larger target model.
To improve verification efficiency, SpecInfer~\cite{miao2024specinfer} introduced token-tree verification to check multiple draft branches simultaneously.
Instead of using a separate model, LayerSkip~\cite{elhoushi2024layerskip} adopted a self-speculative approach, utilizing the target model's own early layers for drafting to reduce memory overhead.

Recent advancements focus on architectures that incorporate auxiliary drafting modules to mitigate the discrepancy between the draft model and the target model.
Medusa~\cite{cai2024medusasimplellminference} adds multiple MLP heads onto the target backbone, enabling the simultaneous prediction of tokens at multiple future positions.
Addressing Medusa's independence assumption, Hydra~\cite{ankner2024hydrasequentiallydependentdraftheads} introduces sequentially dependent heads to capture inter-token correlations.
Eagle~\cite{li2025eaglespeculativesamplingrequires} and HASS~\cite{yang2025hass} use feature-level speculation to align the representation of the draft model with the target model.
Eagle-3~\cite{li2025eagle3scalinginferenceacceleration} further advances this by adopting direct token prediction to unlock scaling laws~\cite{kaplan2020scalinglawsneurallanguage}, while using a training-time test scheme to mitigate error accumulation.

\subsubsection{Model-Free Methods} In contrast, model-free approaches capitalize on the redundancy of context to generate drafts without training additional parameters. Retrieval-based methods, such as REST~\cite{he2024restretrievalbasedspeculativedecoding}, fetch candidates from external or contextual data stores. To exploit internal patterns, Prompt Lookup Decoding (PLD)~\cite{saxena2023prompt} utilizes N-gram repetitions within the prompt. To handle complex structures, Suffix Decoding~\cite{oliaro2025suffixdecodingextremespeculativedecoding} leverages dynamic suffix trees to efficiently retrieve variable-length patterns.
Methods like Lookahead Decoding~\cite{fu2024breaksequentialdependencyllm} and Parallel Decoding~\cite{santilli2023accelerating} break the strict sequential dependency via Jacobi iteration or N-gram pooling, treating the decoding process as solving a system of nonlinear equations in parallel, all without relying on auxiliary models.

Based on this taxonomy, our study selects representative methods from both categories.
This selection establishes a balanced experimental setup, allowing us to rigorously evaluate the performance trade-offs between parametric reasoning and heuristic pattern matching in SE scenarios.


\subsection{Benchmarks for LLMs in SE}
\label{subsec:se_benchmarks}

LLMs show strong capabilities in SE~\cite{hou2024large}, and numerous benchmarks have been proposed to assess their performance on various SE tasks~\cite{hu2025assessing}, such as code generation and understanding~\cite{chen2021evaluatinglargelanguagemodels,austin2021programsynthesislargelanguage,jain2024livecodebench,chen2024code,chen2025reasoning,chen2026securevibebench}, program repair~\cite{yang2024swe,just2014defects4j, yang2026securepair}, and code editing~\cite{aider2024polyglot,cassanocan}.

These scenarios represent common activities in the daily workflows of software developers and engineers. 
In these practical applications, the responsiveness of the model is an important factor that directly influences the efficiency of the software development lifecycle. 
Reducing inference latency can significantly improve the user experience by enabling more fluid and interactive coding assistance. 
Therefore, optimizing the inference speed for these specific tasks is of great practical importance for the widespread adoption of LLMs in the software industry.
In the following, we provide a detailed introduction to the datasets associated with these three usage scenarios:

(i) For code generation, HumanEval~\cite{chen2021evaluatinglargelanguagemodels} and MBPP~\cite{austin2021programsynthesislargelanguage} are two widely used benchmarks for evaluating function-level coding. They require models to complete a standalone function based on a natural language description and correctness is verified using test cases.
To address the issue of data contamination, LiveCodeBench~\cite{jain2024livecodebench} continuously collects new problems from coding contests like LeetCode~\cite{leetcode} and provides a reliable evaluation on competition-level programming. 
At the repository level, benchmarks such as  RepoBench~\cite{liu2023repobenchbenchmarkingrepositorylevelcode} and CrossCodeEval~\cite{ding2023crosscodeevaldiversemultilingualbenchmark} have also been proposed for code generation.
(ii) Program repair focuses on identifying and fixing software defects. 
Benchmarks like Defects4J~\cite{just2014defects4j} and HumanEval-Fix~\cite{muennighoff2024octopackinstructiontuningcode} are widely used to evaluate model performance on identifying and fixing bugs within specific functions or classes.
To represent modern SE, SWE-bench~\cite{jimenez2024swebench} evaluates the ability to resolve real-world GitHub issues, requiring models to understand the repository context and handle long-range dependencies to generate valid patches.
(iii) Code Editing focuses on modifying existing source code to follow user instructions.
These tasks evaluate how well models follow instructions in different settings, ranging from single-turn editing in benchmarks like CanItEdit~\cite{cassanocan} to multi-turn interactions in environments like Aider Polyglot~\cite{aider2024polyglot}, where models must update their code based on test feedback.

These benchmarks are widely used in academia and industry~\cite{grattafiori2024llama,lozhkov2024starcoder2,deepseek2024v2} to present models' capabilities in real-world coding and programming; we therefore follow their choice, aiming for our work to be representative and practically relevant.

\section{Methodology}
\label{sec:methodology}


\subsection{Studied SD Methods}
Our method selection prioritizes representativeness and reproducibility. To ensure a fair evaluation and avoid potential bias from reimplementation, we require all evaluated methods to be officially supported by the same inference framework. Based on these considerations and the taxonomy introduced in \S\ref{subsec:sd_formulation}, we select four SD approaches. These include two model-free methods (Prompt Lookup Decoding and Suffix Decoding) and two model-based methods (MLP Speculators and Eagle-3), which cover both classic techniques and state-of-the-art advancements. These methods accelerate the inference of LLMs with diverse techniques (e.g., structural retrieval, neural-based drafting) and have gained wide adoption in various research domains~\cite{xia2024unlockingefficiencylargelanguage}. In the following, we begin by describing the formulation of SD, and then introduce the selected approaches.

\begin{figure}[t]
    \centering
    \includegraphics[width=\linewidth]{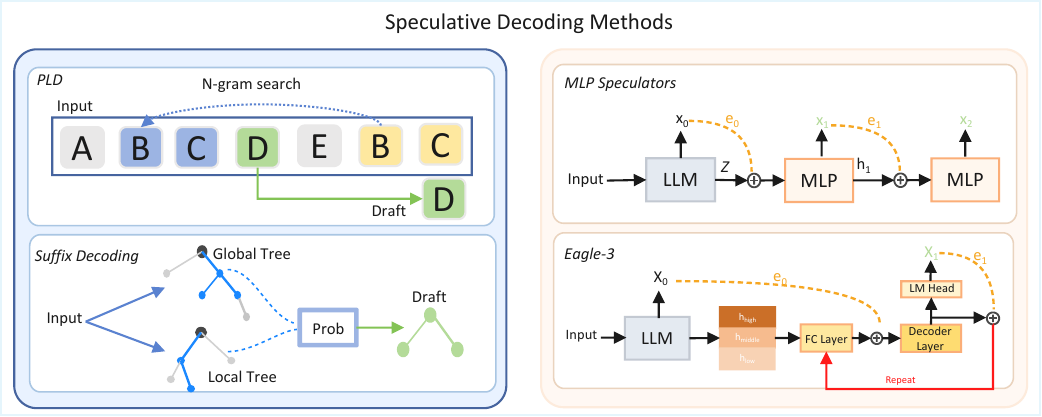}
    \caption{Overview of the studied SD methods.}
    \label{fig:studied_methods}
\end{figure}

\subsubsection{Problem Formulation} 
Let $M_q$ denote the target Large Language Model parameterized by $\theta_q$. Given a historical context sequence $x_{<t}$, the standard autoregressive inference computes the probability distribution $P_{M_q}(x_t | x_{<t})$ for the next token.

To accelerate this process, SD utilizes a speculative drafter, denoted as $M_p$. The objective of $M_p$ is to efficiently synthesize a candidate chain $\hat{x}_{t:t+\gamma}$ of length $\gamma$ (referring to the speculation budget). The goal is to maximize the acceptance rate of these candidates by $M_q$ during the parallel verification phase, which is governed by Greedy Decoding (Eq. \ref{eq:greedy}) or Speculative Sampling, while minimizing the computational overhead of $M_p$ itself. Please refer to \S\ref{subsec:sd_formulation} for the detailed formulation.

\subsubsection{Model-Free Methods}
\label{subsec:model_free}
Model-free methods exploit the intrinsic contextual repetition without additional training overhead.

\phead{Prompt Lookup Decoding (PLD)~\cite{saxena2023prompt}.} 
As illustrated in the upper-left corner of Figure~\ref{fig:studied_methods}, PLD implements a deterministic N-gram retrieval strategy over the entire context. The method operates by matching the current context suffix with previous occurrences in the input sequence. 
As shown in the figure, the input sequence ends with the suffix "B C".
The algorithm identifies a prior occurrence of this pattern and observes that it is followed by the token "D".
Consequently, PLD retrieves "D" as the draft token.
Formally, if a suffix $x_{t-N:t}$ (where $N$ can vary up to a maximum $N_{max}$) appeared previously at position $k$, the method assumes the sequence following that historical occurrence is a likely candidate.
Upon a successful match (visualized as the dotted arrow in the figure), PLD copies a fixed number of subsequent tokens as the speculation:
\begin{equation}
\hat{x}_{t+i} = x_{k+N+i}, \quad \forall i \in \{1, \dots, K\}.
\end{equation}

PLD is primarily constrained by its static speculation budget $K$. This fixed budget fails to adapt to the varying confidence levels of the retrieved context; it may either be too aggressive in uncertain scenarios, leading to frequent verification rollbacks, or too conservative when the historical pattern is highly repetitive and predictable.

\phead{Suffix Decoding~\cite{oliaro2025suffixdecodingextremespeculativedecoding}.}
To overcome the fixed budget limitation of PLD, as visualized in the lower-left of Figure~\ref{fig:studied_methods}, Suffix Decoding utilizes Suffix Trees for probabilistic structural retrieval.
As depicted in the schematic, the method maintains two distinct data structures: a Global Tree that indexes historical outputs from prior requests, and a Local Tree that tracks the current ongoing generation.
The core mechanism involves querying these trees to retrieve frequent patterns.
Nodes in the trees encode the occurrence frequency $\text{Count}(s)$ of sub-sequences.
The ``Prob.'' block in the figure represents the scoring phase, where the conditional probability of a next token $v$ given a suffix context $s$ is estimated via:
\begin{equation}
P_{\text{suffix}}(v | s) = \frac{\text{Count}(s \oplus v)}{\text{Count}(s)},
\end{equation}
where $\oplus$ denotes the concatenation. Based on these probabilities, the method performs a greedy expansion to construct a Speculation Tree (shown as the tree-structured ``Draft'' in the figure).
Moreover, it employs an Adaptive Speculation Strategy: the speculation budget $\gamma$ (the total token count in the draft tree) is dynamically scaled proportional to the pattern match length $L_{\text{match}}$ ($\gamma = \lfloor \alpha \cdot L_{\text{match}} \rfloor$).
This allows the method to perform ``extreme speculation'' by generating deep draft trees when historical patterns are strong, while falling back to conservative drafting for novel logic.

This approach is specifically tailored for agentic applications, particularly automated coding assistants and multi-agent workflows, where workloads exhibit high repetitiveness~\cite{oliaro2025suffixdecodingextremespeculativedecoding}.

\subsubsection{Model-Based Methods}
This category leverages auxiliary trainable parameters to approximate the target distribution.

\phead{MLP Speculators~\cite{wertheimer2024acceleratingproductionllmscombined}.}
We utilize the Multi-stage MLP architecture proposed by Wertheimer et al.~\cite{wertheimer2024acceleratingproductionllmscombined}, as illustrated in Figure~\ref{fig:studied_methods}.
This method attaches a series of lightweight Multi-Layer Perceptron (MLP) layers to the top of the frozen target LLM.
Instead of predicting multiple future tokens independently, it employs a multi-stage generation approach with sequential dependencies.
Specifically, as visually depicted by the sequential chain in the figure, the prediction of the $k$-th future token depends on the hidden state of the $(k-1)$-th step. This allows the drafter to capture the logical dependencies between consecutive tokens (i.e., the second word depends on the first word) while maintaining high inference speed.

Formally, let $z$ denote the final hidden state of the base model (Stage 0). The state $h_k$ for the $k$-th speculative step is derived recursively:
\begin{equation}
    h_{k} = \text{LN}(\text{Act}(\text{Proj}(h_{k-1} \oplus e_{k-1}))), \quad P_{\text{spec}}(x_{t+k}) = \text{Softmax}(W^{(k)}_{\text{head}} \cdot h_k),
\end{equation}
where $h_0 = z$, $e_{k-1}$ is the embedding of the sampled token from the previous stage, and $\oplus$ denotes concatenation.
$\text{Proj}_k$ and $W_{\text{head}}^{(k)}$ refer to the linear projection and prediction head specific to stage $k$; notably, these layers do not share weights across stages to maximize adaptability. Finally, $\text{Act}$, $\text{LN}$, and $\text{Softmax}$ correspond to the GeLU activation, Layer Normalization, and standard softmax function, respectively.
While introducing sequential dependencies significantly enhances the quality of generated drafts and their alignment with the target LLM, it necessitates a complex two-stage training regime to effectively align the speculator's output distribution with that of the target LLM.

\phead{Eagle-3~\cite{li2025eagle3scalinginferenceacceleration}.}
The Eagle series~\cite{li2024eagle2fasterinferencelanguage,li2025eaglespeculativesamplingrequires,li2025eagle3scalinginferenceacceleration} is a prominent family of SD frameworks.
Specifically, we select Eagle-3~\cite{li2025eagle3scalinginferenceacceleration}, illustrated in lower right Figure~\ref{fig:studied_methods}, to represent state-of-the-art model-based speculation.
Unlike its predecessor (Eagle~\cite{li2025eaglespeculativesamplingrequires}) which relies on feature-level regression, Eagle-3 shifts to direct token prediction and introduces TTT alongside Multi-Layer Feature Fusion to effectively leverage scaling laws~\cite{kaplan2020scalinglawsneurallanguage}.
As visually detailed in the figure, instead of relying solely on top-layer features, the drafter integrates low-, middle-, and high-level features (labeled as $h_{low}, h_{middle}, h_{high}$) from the target model to capture richer semantic information.
TTT does not imply updating parameters during inference; rather, it refers to a novel training paradigm where the drafter simulates the multi-step generation process during training.
While removing the feature prediction constraint enhances the model's expressiveness and scalability, it creates a discrepancy between training inputs (ground truth) and testing inputs (predicted vectors).
TTT addresses this by exposing the model to its own predicted deviations during training, thereby mitigating the distribution shift typically encountered during autoregressive generation.
Formally, let $g_t$ be the fused features from the target model and $e_t$ be the token embedding. The drafter generates an unconstrained vector $a_{t+1}$, which is then projected by the target model's LM head to predict the next token:
\begin{equation}
a_{t+1} = \text{Transformer}_{\theta_{\text{draft}}}(e_t, g_t), \quad P_{\text{draft}}(x_{t+1}) = \text{Softmax}(W_{\text{target\_head}} \cdot a_{t+1})
\end{equation}

\subsection{Evaluation Metrics}
Following previous research~\cite{li2025eagle3scalinginferenceacceleration,oliaro2025suffixdecodingextremespeculativedecoding,chen2023accelerating,saxena2023prompt}, we employ three key metrics to quantify the acceleration performance of SD. For all metrics listed below, higher values indicate better performance.
\begin{itemize}
    \item \textbf{Speedup} is defined as the ratio of the generation throughput (i.e., the number of tokens produced per second) of the speculative method to that of the autoregressive baseline.
    \item \textbf{Mean Acceptance Length ($\tau$)} is the expected number of candidate tokens validated by the target model in a single verification step. This metric serves as a direct proxy for the drafting model's quality and the semantic alignment between the drafter and the verifier.
    \item \textbf{Acceptance Rate ($n-\alpha$)} denotes the acceptance rate $\alpha$ at a specific position index $n$ within the draft chain. It is calculated as the conditional probability that the draft token at position $n$ is accepted, given that all preceding tokens ($0 \dots n-1$) have been verified as correct. For example, we calculate $3-\alpha$ by dividing the number of tokens accepted at position 3 by the number of tokens accepted at position 2.
\end{itemize}

\subsection{SE Benchmarks}
We evaluate SD on six representative datasets covering three common SE scenarios: code generation, program repair, and code editing. For each scenario, we select two benchmarks to enrich the benchmark suite and support a more comprehensive evaluation.

\begin{itemize}
    \item \textbf{LiveCodeBench~\cite{jain2024livecodebench}}: 
    LiveCodeBench is a benchmark designed to evaluate the code generation capabilities of LLMs on competition-level programming problems. It collects problems from contests such as LeetCode, AtCoder, and Codeforces.
    The task involves generating a standalone solution function based on a natural language problem description and passing a comprehensive set of test cases.
    For our experiments, we utilize the \texttt{release\_v6} dataset comprising 1,055 problems. 
    We bypass the standard temporal filtering (designed to prevent data contamination), since our focus is on measuring decoding latency rather than evaluating  generalization.

    \item \textbf{HumanEval~\cite{chen2021evaluatinglargelanguagemodels}}:
    HumanEval is a widely used benchmark for evaluating function-level code generation.
    Each task requires the model to synthesize a complete Python function from a natural language specification, and correctness is checked by executing unit tests against the generated solution.
    In our experiments, we use the standard HumanEval split with 164 problems.

    \item \textbf{SWE-bench Verified~\cite{jimenez2024swebench} (SWE-bench)}: 
    SWE-bench evaluates the ability of LLMs and autonomous agents to resolve real-world SE issues collected from popular GitHub repositories.
    Given a full codebase and a natural language issue description, the model is required to navigate the repository context and generate a patch file to fix the reported bug.
    In this study, we employ the \texttt{Verified} split (comprising 500 issues), which has undergone manual quality assurance to ensure solvability. 

    \item \textbf{Defects4J~\cite{just2014defects4j}}:
    Defects4J is a classic benchmark for automated program repair on real Java bugs.
    We evaluate it through the D4C~\cite{xu2024aligning} framework, which formulates repair as direct debugging rather than first localizing buggy hunks and then applying an infilling-style patch.
    Specifically, D4C constructs prompts from the buggy function, its comment, failed JUnit test, and error message, and asks the model to regenerate the repaired function.

    \item \textbf{Aider Polyglot~\cite{aider2024polyglot}}: 
    The Aider Polyglot benchmark assesses the code editing capabilities of models across diverse programming languages within a chat-based interface.
    The task simulates a pair programming scenario: the model is presented with an existing source file and a user instruction to modify the logic.
    We utilize the polyglot-benchmark suite (comprising 225 tasks). We standardize the interaction on the \texttt{whole} edit format. We adopt this format following the Aider developers' recommendation to ensure robustness for models with limited diff-generation capabilities, requiring the model to regenerate the entire source file.

    \item \textbf{CanItEdit~\cite{cassanocan}}:
    CanItEdit is a code editing benchmark that measures whether models can follow natural language instructions to revise existing source code.
    The tasks are designed around localized edits, where the model must update a given program to satisfy a requested behavior change while preserving the surrounding implementation.
    In our experiments, we use the full benchmark suite to evaluate edit quality and decoding efficiency in instruction-driven modification scenarios.
\end{itemize}

\subsection{Subject LLMs}

We employ four widely used open-weights models selected to represent a diverse spectrum of parameter scales, activation patterns, and reasoning capabilities, spanning from lightweight architectures suitable for edge deployment to large-scale models approximating frontier-level performance.
Their corresponding model IDs on Hugging Face\footnote{\url{https://huggingface.co/}} are listed as follows:

\begin{itemize}

    \item \textbf{Qwen/Qwen3-32B~\cite{qwen3technicalreport}(Qwen3-32B)}:
    Qwen3-32B is the primary dense model of the Qwen3 family, designed to bridge the gap between varying inference budgets through a novel Hybrid Thinking architecture.
    Released by the Qwen Team in April 2025, it features 32.8 billion parameters and supports dynamic switching between a compute-intensive ``Thinking Mode'' (for complex reasoning) and a latency-optimized ``Non-Thinking Mode''.
    For this study, we explicitly disable the Thinking Mode (setting \texttt{enable\_thinking=False}).

    \item \textbf{Qwen/Qwen3-30B-A3B-Instruct-2507~\cite{qwen3technicalreport}(Qwen3-30A3)}:
    Qwen3-30A3 is a mixture-of-experts variant in the Qwen3 family, designed to improve the trade-off between model capacity and inference efficiency through sparse activation.
    It contains 30.5 billion total parameters while activating 3.3 billion parameters per token, enabling it to retain broad instruction-following and coding capabilities with lower per-token computational cost.
    As an instruction-only variant, it supports the latency-optimized non-thinking mode by default and does not require explicitly disabling thinking behavior during inference.
    
    \item \textbf{meta-llama/Llama-3.1-8B-Instruct~\cite{grattafiori2024llama}(L31-8B)}: 
    L31-8B is the lightweight entry in the Llama 3.1 series, designed to deliver strong reasoning capabilities within a compact parameter budget suitable for edge deployment.
    Developed by Meta, it utilizes a standard dense transformer architecture with a 128k context window and is optimized for multilingual dialogue and code generation tasks.

    \item \textbf{meta-llama/Llama-3.1-70B-Instruct~\cite{grattafiori2024llama}(L31-70B)}: 
    L31-70B represents a tier with high performance in the Llama 3.1 series, positioned to approximate the capabilities of frontier proprietary models in long-context understanding and code synthesis.
    It utilizes a large-scale dense transformer architecture trained on over 15 trillion tokens, excelling in complex instruction following and reasoning tasks.
\end{itemize}

\subsection{Implementation Details}
\label{subsec:details}

\phead{Experimental Environments.} All experiments are conducted on a Linux server with 6 NVIDIA A800 (80GB) GPUs.
The system is powered by \textit{Intel Xeon Platinum 8358P} processors and operates on \textit{Ubuntu 20.04.6 LTS}.

\phead{Studied Models.} 
We use a popular LLM inference library vLLM~\cite{kwon2023efficient} v0.12.0 with 8 concurrent threads to implement the local deployment. 
We standardize the maximum context length to 32,768 across all models and limit generation to 1,024 tokens per turn for agent interactions.
For Qwen3-32B, we explicitly disable its thinking mode by setting the parameter \texttt{enable\_thinking} to \texttt{False}. 
For issue-solving tasks in SWE-bench, we adopt \texttt{mini-swe-agent}~\cite{yang2024swe}, a popular and powerful framework as the agent scaffold.

\phead{SD Approaches.} To ensure reproducibility and prioritize precision in code generation, we employ greedy decoding across all experiments. 
For detailed configurations of different methods, we conduct the experiments following their common practice: 
(i) For PLD, it is configured with a lookup window size $N=4$ and a draft length $K=5$, adhering to the default recommendations provided by the vLLM framework. 
(ii) For Suffix Decoding, we use a maximum draft length of $K=32$, adhering to the authors' recommended default settings~\cite{oliaro2025suffixdecodingextremespeculativedecoding}. 
(iii) For Eagle-3, the speculative length is fixed at 3. We use official RedHatAI\footnote{\url{https://huggingface.co/RedHatAI}} checkpoints for 8B/30B/32B models; for the 70B model, due to the unavailability of pre-trained drafter weights, we train a custom drafter via SpecForge~\cite{specforge2025} on BAAI/TACO~\cite{li2023taco} and SWE-Gym/SWE-Gym~\cite{pan2025training} with a learning rate of $5\text{e-}5$. 
(iv) For MLP Speculator, we employ the standard implementation with default hyperparameters, utilizing the pre-trained draft models provided by IBM AI Platform\footnote{\url{https://huggingface.co/ibm-ai-platform}}. Note that this baseline is evaluated on vLLM v0.9.2 due to specific version compatibility constraints, which restrict it to single-threaded execution.

\section{Results}
\label{sec:evaluation}

In this section, we present the results and analysis to answer the
research questions:
\begin{itemize}
    \item \textbf{RQ1: How effective is SD on various SE tasks?}

    \item \textbf{RQ2: How do model scale and architecture impact acceleration?}

    \item \textbf{RQ3: How does the performance of SD differ between natural language and code?}

\end{itemize}

\subsection{RQ1: Overall Effectiveness of SD}
\label{subsec:rq1}

\begin{table*}[t]
    \centering
    \caption{Performance comparison of SD methods across different benchmarks. We report Speedup relative to the autoregressive decoding baseline (1.00$\times$) and Mean Acceptance Length ($\tau$). We omit $\tau$ for MLP Speculators due to framework version limitations. \textbf{Bold} indicates the best performance in each category.}
    \label{tab:main_results}
    \resizebox{\textwidth}{!}{%
    \begin{tabular}{llcccccccccccccc}
        \toprule
        \multirow{2}{*}{\textbf{Model}} & \multirow{2}{*}{\textbf{Method}} & \multicolumn{2}{c}{\textbf{LiveCodeBench}} & \multicolumn{2}{c}{\textbf{HumanEval}} & \multicolumn{2}{c}{\textbf{SWE-bench}} & \multicolumn{2}{c}{\textbf{Defects4J}} & \multicolumn{2}{c}{\textbf{Aider Polyglot}} & \multicolumn{2}{c}{\textbf{CanItEdit}} & \multicolumn{2}{c}{\textbf{Mean}} \\
        \cmidrule(lr){3-4} \cmidrule(lr){5-6} \cmidrule(lr){7-8} \cmidrule(lr){9-10} \cmidrule(lr){11-12} \cmidrule(lr){13-14} \cmidrule(lr){15-16}
         & & \textbf{Speedup} & $\tau$ & \textbf{Speedup} & $\tau$ & \textbf{Speedup} & $\tau$ & \textbf{Speedup} & $\tau$ & \textbf{Speedup} & $\tau$ & \textbf{Speedup} & $\tau$ & \textbf{Speedup} & $\tau$ \\
        \midrule

        \multirow{3}{*}{Qwen3-32B}
          & PLD & 1.13$\times$ & \textbf{2.89} & 1.17$\times$ & \textbf{2.89} & 1.05$\times$ & 2.92 & 1.87$\times$ & \textbf{3.40} & 1.08$\times$ & \textbf{3.19} & 2.23$\times$ & 4.07 & 1.42$\times$ & \textbf{3.23} \\
          & Suffix Decoding & 1.27$\times$ & 1.21 & 1.56$\times$ & 1.10 & \textbf{1.66$\times$} & \textbf{3.48} & \textbf{2.61$\times$} & 3.33 & 1.22$\times$ & 1.35 & \textbf{3.05$\times$} & \textbf{4.11} & \textbf{1.90$\times$} & 2.43 \\
          & Eagle-3 & \textbf{1.71$\times$} & 1.87 & \textbf{1.94$\times$} & 1.99 & 1.10$\times$ & 1.87 & 1.21$\times$ & 2.10 & \textbf{1.29$\times$} & 1.81 & 2.14$\times$ & 2.35 & 1.57$\times$ & 2.00 \\
        \midrule

        \multirow{3}{*}{Qwen3-30A3}
          & PLD & 0.87$\times$ & \textbf{2.21} & 0.99$\times$ & \textbf{2.57} & 1.02$\times$ & 3.25 & 1.36$\times$ & \textbf{3.38} & 1.08$\times$ & \textbf{3.29} & 1.55$\times$ & \textbf{3.83} & 1.15$\times$ & \textbf{3.09} \\
          & Suffix Decoding & \textbf{0.95$\times$} & 1.02 & \textbf{1.03$\times$} & 0.91 & \textbf{1.73$\times$} & \textbf{4.78} & \textbf{1.53$\times$} & 3.14 & 0.99$\times$ & 1.41 & \textbf{1.83$\times$} & 2.93 & \textbf{1.34$\times$} & 2.37 \\
          & Eagle-3 & 0.77$\times$ & 1.75 & 0.54$\times$ & 1.93 & 0.41$\times$ & 0.43 & 0.55$\times$ & 1.57 & \textbf{1.15$\times$} & 1.52 & 0.87$\times$ & 2.22 & 0.72$\times$ & 1.57 \\
        \midrule

        \multirow{3}{*}{L31-8B}
          & PLD & 1.07$\times$ & \textbf{2.36} & 1.16$\times$ & \textbf{2.81} & 1.71$\times$ & 2.99 & 1.78$\times$ & \textbf{3.23} & 1.30$\times$ & \textbf{3.36} & 2.05$\times$ & \textbf{4.01} & 1.51$\times$ & \textbf{3.13} \\
          & Suffix Decoding & 1.58$\times$ & 1.30 & 1.63$\times$ & 1.13 & \textbf{3.35$\times$} & \textbf{5.48} & \textbf{2.74$\times$} & 3.18 & \textbf{1.82$\times$} & 1.99 & \textbf{2.90$\times$} & 3.54 & \textbf{2.34$\times$} & 2.77 \\
          & Eagle-3 & \textbf{1.77$\times$} & 1.82 & \textbf{2.19$\times$} & 2.11 & 0.70$\times$ & 0.64 & 1.55$\times$ & 1.56 & 1.32$\times$ & 1.49 & 2.12$\times$ & 2.18 & 1.61$\times$ & 1.63 \\
        \midrule

        \multirow{4}{*}{L31-70B}
          & PLD & 0.97$\times$ & \textbf{2.39} & 0.99$\times$ & \textbf{2.40} & 1.02$\times$ & 2.69 & 1.67$\times$ & \textbf{3.77} & 1.12$\times$ & \textbf{3.40} & 1.71$\times$ & \textbf{4.05} & 1.25$\times$ & \textbf{3.12} \\
          & Suffix Decoding & 1.24$\times$ & 1.08 & 1.13$\times$ & 1.00 & \textbf{1.79$\times$} & \textbf{3.39} & \textbf{1.70$\times$} & 3.55 & \textbf{1.50$\times$} & 2.72 & \textbf{1.99$\times$} & 3.93 & \textbf{1.56$\times$} & 2.61 \\
          & Eagle-3 & \textbf{1.66$\times$} & 1.74 & \textbf{1.41$\times$} & 1.27 & 0.87$\times$ & 1.07 & 0.83$\times$ & 0.35 & 0.88$\times$ & 0.48 & 1.29$\times$ & 1.09 & 1.16$\times$ & 1.00 \\
          & MLP Speculators & 0.81$\times$ & --- & 1.22$\times$ & --- & 0.91$\times$ & --- & 0.88$\times$ & --- & 1.23$\times$ & --- & 1.03$\times$ & --- & 1.01$\times$ & --- \\

        \bottomrule
    \end{tabular}%
    }
\end{table*}

\phead{General Effectiveness.}
As presented in Table~\ref{tab:main_results}, our evaluation shows that SD is an effective technique for accelerating LLM inference in SE tasks. Notably, the mean speedups of most model–method combinations exceed 1.00$\times$, demonstrating the general effectiveness of this technique in reducing inference latency. However, the actual performance exhibits high variance across different setups. 
Specifically, excluding MLP Speculators due to its different inference setup, the unweighted mean speedup across all PLD, Suffix Decoding, and Eagle-3 results in Table~\ref{tab:main_results} is 1.46$\times$.
On one end of the spectrum, Suffix Decoding achieves a 3.35$\times$ speedup on L31-8B for SWE-bench. Conversely, we also observe significant instability, with performance dropping to a 0.41$\times$ slowdown in the worst-case scenario (Eagle-3 with Qwen3-30A3 on SWE-bench). 
This data point reveals that while SD can accelerate LLM inference for code tasks, it may also have a negative effect in some cases. 
We discuss the MLP Speculators method separately due to potential bias introduced by the inference infrastructure (see Section~\ref{subsec:details} for details).
As shown in Table~\ref{tab:main_results}, this method achieves a notable 1.23$\times$ speedup on Aider Polyglot but exhibits limited acceleration overall, resulting in an average speedup of 1.01$\times$ for L31-70B across the six benchmarks. We attribute this outcome to the advanced kernel optimizations~\cite{vllm_v0120_release} in vLLM, which accelerate the baseline and consequently limit the realized speedup.

\begin{findingbox}
\textbf{Finding 1:}
SD shows clear acceleration potential on SE tasks, with an average speedup of 1.46$\times$ across six benchmarks and four subject models.
However, the gains are highly configuration-dependent, and some settings even lead to slower inference.
\end{findingbox}

\begin{figure*}[h]
    \centering
    \begin{subfigure}[b]{0.32\textwidth}
        \centering
        \includegraphics[width=\linewidth]{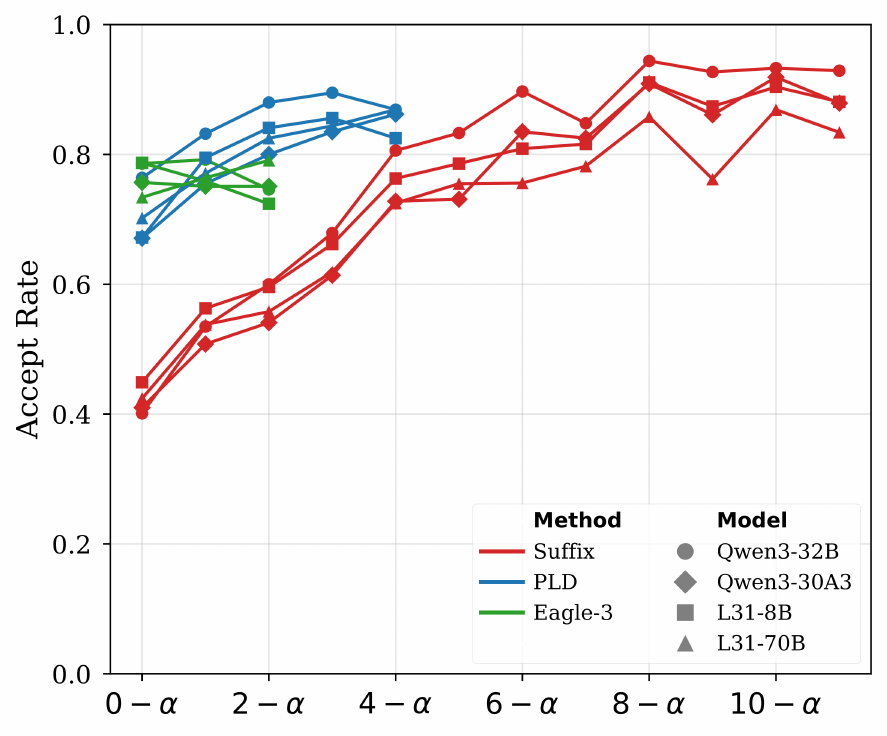}
        \caption{\textbf{LiveCodeBench}}
        \label{fig:n_alpha_lcb}
    \end{subfigure}
    \hfill 
    \begin{subfigure}[b]{0.32\textwidth}
        \centering
        \includegraphics[width=\linewidth]{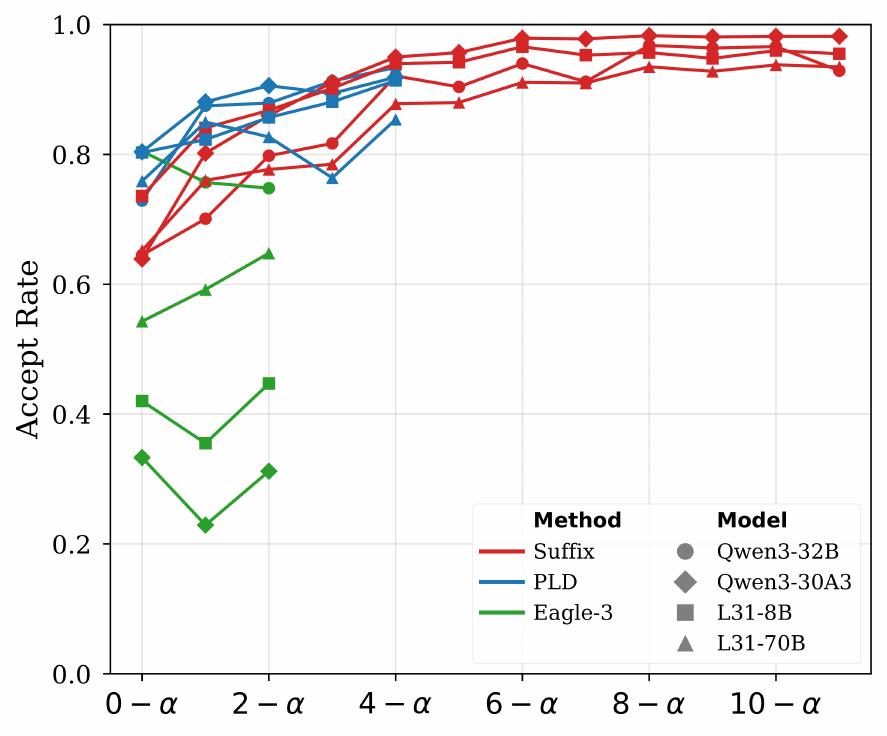}
        \caption{\textbf{SWE-bench}} 
        \label{fig:n_alpha_swe}      
    \end{subfigure}
    \hfill 
    \begin{subfigure}[b]{0.32\textwidth}
        \centering
        \includegraphics[width=\linewidth]{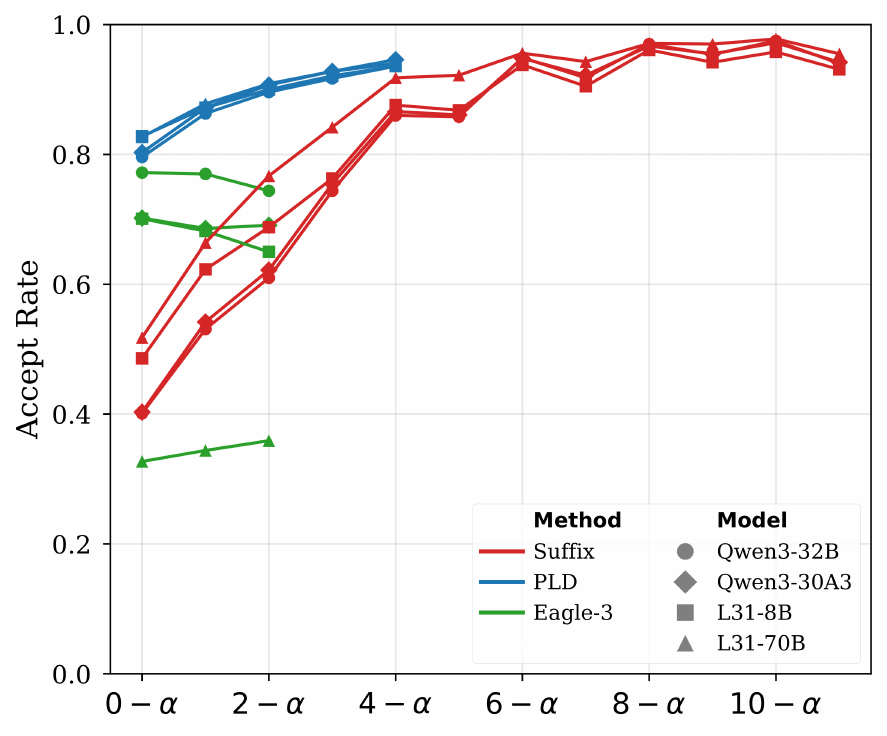}
        \caption{\textbf{Aider Polyglot}} 
        \label{fig:n_alpha_aider}         
    \end{subfigure}
    
    \caption{The $n-\alpha$ curves on (a) LiveCodeBench, (b) SWE-bench, and (c) Aider Polyglot.}
    \label{fig:n_alpha_all}
\end{figure*}

\phead{Suitability of SD Methods across Task Categories.}
As detailed in Table~\ref{tab:main_results}, our empirical evaluation reveals that no single SD method serves as a universally optimal solution; rather, the performance of SD varies across SE task categories. Figure~\ref{fig:n_alpha_all} further illustrates this trend on three representative benchmarks, one from each category.

For code generation, represented by LiveCodeBench and HumanEval, Eagle-3 demonstrates superior performance. As shown in Table~\ref{tab:main_results}, Eagle-3 achieves an average speedup of 1.50$\times$ across these two benchmarks, outperforming Suffix Decoding (1.30$\times$) and PLD (1.04$\times$).
We attribute this advantage to the characteristic of the task, which involves translating natural language into code; 
since the ground-truth code does not appear in the context history, model-free methods fail to find sufficient repetitive patterns to copy.
As shown in Table~\ref{tab:main_results}, Suffix Decoding achieves a mean $\tau$ of 1.09 on code generation benchmarks, compared to 3.79 on code repair benchmarks and 2.75 on code editing benchmarks, respectively. PLD exhibits a similar trend, recording a mean $\tau$ of 2.57 on code generation benchmarks, which is consistently lower than the 3.20 and 3.65 observed on code repair and code editing benchmarks.

In contrast, for code repair and code editing tasks, Suffix Decoding demonstrates distinct superiority. Across SWE-bench and Defects4J, Suffix Decoding achieves an average speedup of 2.14$\times$, compared with 1.44$\times$ for PLD and 0.90$\times$ for Eagle-3. Across Aider Polyglot and CanItEdit, Suffix Decoding reaches 1.91$\times$, compared with 1.52$\times$ for PLD and 1.38$\times$ for Eagle-3.
We attribute this dominance to the higher contextual redundancy in repair and editing scenarios, which is especially pronounced in repository-level agentic workflows such as SWE-bench and also appears when models modify code around existing context.
As suggested by Oliaro et al.~\cite{oliaro2025suffixdecodingextremespeculativedecoding}, such redundancy inherently favors Suffix Decoding, which retrieves explicit repetitive patterns rather than approximating distributions.

For PLD, as detailed in Table~\ref{tab:main_results}, it achieves the highest Mean Acceptance Length ($\tau$) in the majority of evaluated benchmarks. This observation is consistent with Figure~\ref{fig:n_alpha_all}, which illustrates that PLD maintains higher acceptance rates compared to others. However, the method does not translate these favorable metrics into the strongest acceleration, yielding an average speedup of 1.33$\times$ across the evaluated benchmarks. We attribute this discrepancy to its algorithmic complexity. PLD's N-gram retrieval entails a complexity of $O(T)$, where $T$ denotes the current context length. Consequently, the overhead becomes non-negligible, effectively neutralizing the latency gains derived from a higher $\tau$.

\begin{findingbox}
\textbf{Finding 2:} Eagle-3 excels in code generation tasks, whereas Suffix Decoding demonstrates superior performance in code repair and code editing scenarios.
\end{findingbox}

\phead{Diminishing Returns in Long-Context Tasks.}
In SWE-bench, we observe a phenomenon of diminishing returns. Specifically, the increase in speedup slows down even as the mean acceptance length ($\tau$) continues to grow.
We demonstrate this by comparing representative macro-average values of speedup and $\tau$ on LiveCodeBench and SWE-bench, which correspond to code generation and repository-level repair tasks, respectively.
For Suffix Decoding, as shown in Table~\ref{tab:main_results}, although high redundancy in the repository causes $\tau$ to increase from 1.15 to 4.28, the speedup only increases from $1.26\times$ to $2.13\times$. 
Similarly, for Eagle-3 and PLD, we observe that the speedup decreases even when the mean acceptance length ($\tau$) remains stable. Taking Qwen3-32B as an example, Eagle-3 maintains an identical $\tau$ of 1.87 on both benchmarks. However, the speedup drops from 1.71$\times$ to 1.10$\times$. In the same way, for PLD, while $\tau$ stays at a high level (2.89 on LiveCodeBench and 2.92 on SWE-bench), the speedup declines from 1.13$\times$ to 1.05$\times$.
This indicates that draft quality (implied by $\tau$) is not the only factor for acceleration in complex tasks.

We attribute this phenomenon to the extensive context windows of repository-level tasks.
According to Yang et al.~\cite{yang2025longspeclongcontextlosslessspeculative}, in long-context scenarios, the inefficiencies in attention mechanisms and massive KV cache demands significantly impede verification efficiency, offsetting potential speedup offered by the high acceptance rates.
Our analysis of L31-8B request logs reveals that SWE-bench tasks average 15,999 total tokens per request, substantially higher than the 817 tokens for LiveCodeBench. Here, total tokens include both prompt and generated tokens, and the average is computed over assistant model requests. This implies that in repository-level tasks, the verification process transitions into a compute-bound regime, ultimately limiting the realizable speedup.

\begin{findingbox}
\textbf{Finding 3:} In repository-scale tasks, high mean acceptance length ($\tau$) does not translate linearly into speedup gains due to the increased verification overhead.
\end{findingbox}


\phead{Generalization Challenges of Draft Models.}
The effectiveness of Eagle-3 varies substantially across target models. For L31-8B, Qwen3-32B, and Qwen3-30A3, whose draft models are obtained from RedHatAI, the average acceptance length differs markedly: 1.63, 2.00, and 1.57, respectively. This suggests that model-based methods are sensitive to draft-target alignment, which is closely tied to target-model characteristics such as model scale and architecture.

\begin{table*}[t]
    \centering
    \caption{Comparison between the official RedHatAI Eagle-3 drafter and our SE-trained Eagle-3 drafter on Qwen3-32B. The SE-trained drafter uses the same training datasets and experimental setup as the custom L31-70B drafter.}
    \label{tab:qwen32_eagle_training_data}
    \resizebox{\textwidth}{!}{%
    \begin{tabular}{lcccccccccccccc}
        \toprule
        \multirow{2}{*}{\textbf{Drafter Source}} & \multicolumn{2}{c}{\textbf{LiveCodeBench}} & \multicolumn{2}{c}{\textbf{HumanEval}} & \multicolumn{2}{c}{\textbf{SWE-bench}} & \multicolumn{2}{c}{\textbf{Defects4J}} & \multicolumn{2}{c}{\textbf{Aider Polyglot}} & \multicolumn{2}{c}{\textbf{CanItEdit}} & \multicolumn{2}{c}{\textbf{Mean}} \\
        \cmidrule(lr){2-3} \cmidrule(lr){4-5} \cmidrule(lr){6-7} \cmidrule(lr){8-9} \cmidrule(lr){10-11} \cmidrule(lr){12-13} \cmidrule(lr){14-15}
         & \textbf{Speedup} & $\tau$ & \textbf{Speedup} & $\tau$ & \textbf{Speedup} & $\tau$ & \textbf{Speedup} & $\tau$ & \textbf{Speedup} & $\tau$ & \textbf{Speedup} & $\tau$ & \textbf{Speedup} & $\tau$ \\
        \midrule
        RedHatAI & 1.71$\times$ & 1.87 & 1.94$\times$ & 1.99 & 1.10$\times$ & 1.87 & 1.21$\times$ & 2.10 & 1.29$\times$ & 1.81 & 2.14$\times$ & 2.35 & 1.57$\times$ & 2.00 \\
        SE-trained (ours) & 1.34$\times$ & 1.15 & 1.08$\times$ & 0.51 & 0.65$\times$ & 0.60 & 0.71$\times$ & 0.23 & 0.76$\times$ & 0.29 & 1.11$\times$ & 0.65 & 0.94$\times$ & 0.57 \\
        \bottomrule
    \end{tabular}%
    }
\end{table*}

To further separate the effect of target-model characteristics from the effect of drafter training data, we conduct a controlled comparison on Qwen3-32B. Specifically, we train an additional Eagle-3 drafter using the same SE datasets and experimental setup as the custom L31-70B drafter, and compare it with the official RedHatAI drafter on the same target model. As shown in Table~\ref{tab:qwen32_eagle_training_data}, the SE-trained drafter underperforms the RedHatAI drafter across all six benchmarks. Its mean speedup decreases from 1.57$\times$ to 0.94$\times$, and its mean acceptance length drops from 2.00 to 0.57.

This result suggests that domain relevance alone is insufficient for robust model-based speculation. Although the SE-trained drafter uses code- and repository-oriented data, the training corpus may still be too limited in scale and diversity to cover the heterogeneous distributions of SE tasks. In contrast, the RedHatAI drafter achieves stronger generalization on Qwen3-32B, indicating that training data scale, coverage, and training quality can be as important as domain match when building model-based drafters. A similar pattern also appears in our custom L31-70B drafter, which performs better on code generation benchmarks but degrades on several repair and editing benchmarks.


%

\begin{findingbox}
\textbf{Finding 4:} Model-based methods are sensitive to draft-target alignment. To enable the drafter to generalize across downstream task distributions, draft-target alignment requires training data with sufficient scale and coverage over diverse SE workflows.
\end{findingbox}

\subsection{RQ2: The Impact of Model Scale and Architecture}
\label{subsec:rq2}

\phead{Effects of Model Scale and Architecture.}
To investigate the impact of target models, we calculate the grand mean speedup of PLD, Suffix Decoding, and Eagle-3 across the six SE benchmarks for each model.
The results show that SD acceleration varies substantially across target models, suggesting that model scale and architecture both matter.
Among dense models, the average speedup decreases from L31-8B (1.82$\times$) to Qwen3-32B (1.63$\times$) and L31-70B (1.32$\times$), indicating that larger dense models tend to receive smaller relative latency reductions from SD.

This scale-related trend is particularly visible for model-free methods.
For example, Suffix Decoding achieves 2.34$\times$ on L31-8B, 1.90$\times$ on Qwen3-32B, and 1.56$\times$ on L31-70B; PLD similarly decreases from 1.51$\times$ to 1.42$\times$ and 1.25$\times$.
One possible explanation is that verification overhead and baseline decoding efficiency change with model scale, reducing the relative gain available to SD on larger dense models.

However, architecture also matters.
Qwen3-30A3, a sparsely activated MoE model, achieves the lowest average speedup at 1.07$\times$ despite having a total parameter count comparable to Qwen3-32B.
Its Eagle-3 mean acceptance length is 1.57, yet the realized speedup is only 0.72$\times$, suggesting that mean acceptance length alone cannot explain latency gains on this MoE model; the lower inference cost of sparse activation may reduce the additional speedup that SD can provide.

\begin{findingbox}
\textbf{Finding 5:} SD acceleration varies with target-model scale and architecture: larger dense models tend to receive smaller relative speedups, while the sparsely activated MoE model shows limited latency gains due to efficient target inference.
\end{findingbox}

\begin{figure}[t] 
    \centering
    \begin{subfigure}[b]{0.48\linewidth}
        \centering
        \includegraphics[width=\linewidth]{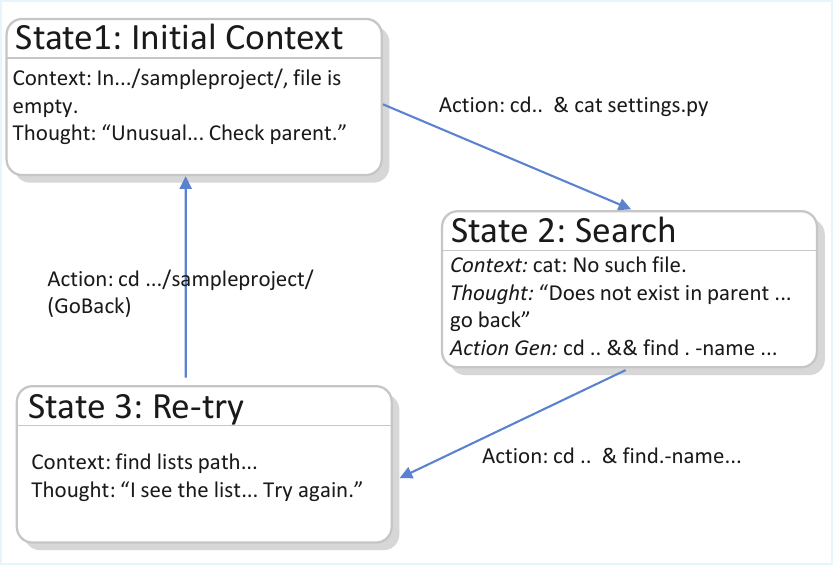}
        \caption{Agentic Loop Process} 
        \label{fig:agentic_loop}
    \end{subfigure}
    \hfill 
    \begin{subfigure}[b]{0.48\linewidth}
        \centering
        \includegraphics[width=\linewidth]{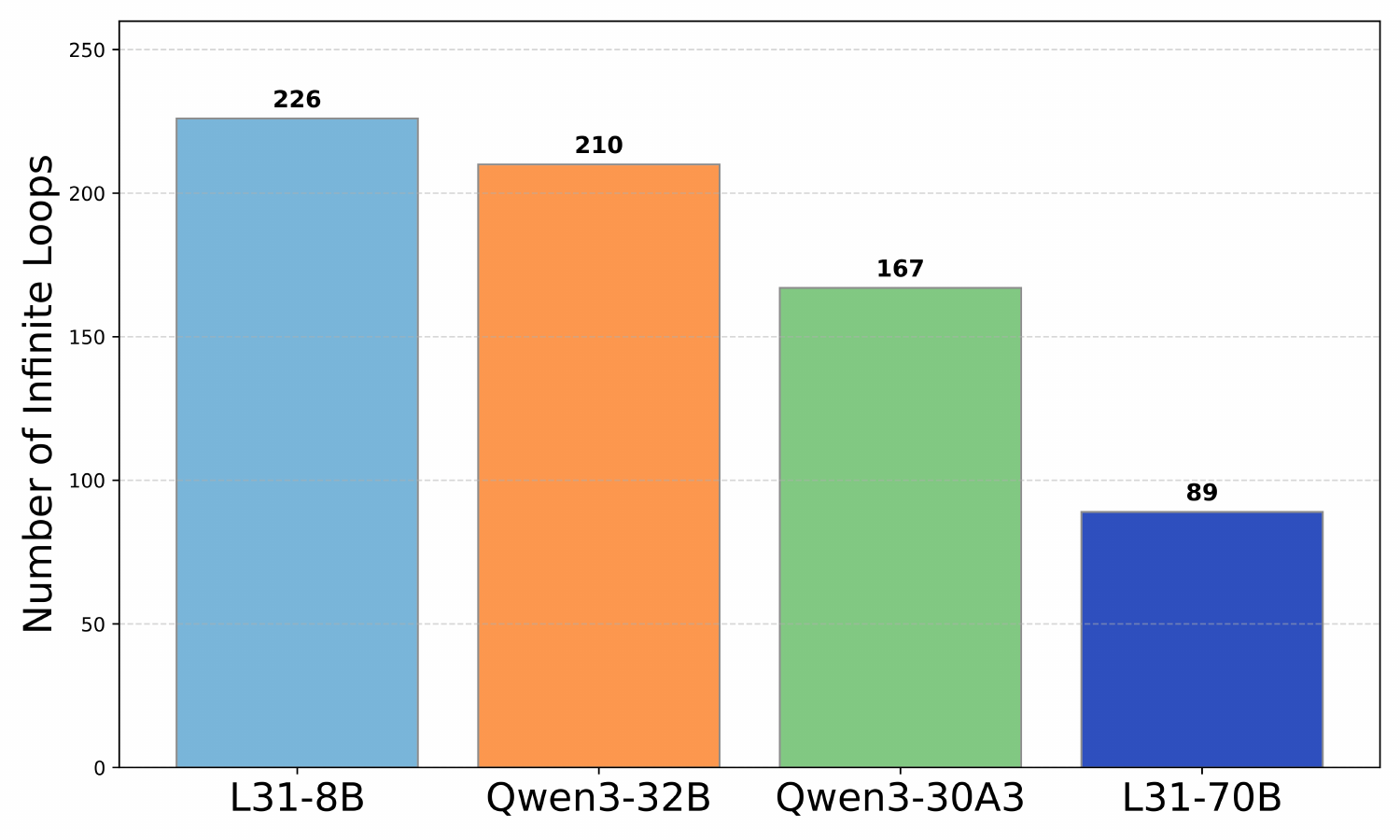}
        \caption{Frequency of Infinite Loops} 
        \label{fig:num_of_loops}
    \end{subfigure}
    
    \caption{Illustration of the agentic loop (a) and the statistics of infinite loops observed across different models (b).}
    \label{fig:main_comparison}
    \vspace{-1.35em}
\end{figure}

\phead{Evaluation Noise from Agentic Infinite Loops.}
In addition to model scale and architecture, agentic infinite loops can introduce additional noise into SD acceleration measurements.
In SWE-bench, we observe that some models enter repetitive action loops, producing nearly identical Thought and Action sequences over multiple turns.
Such loops create artificial context redundancy, which can be exploited by model-free SD methods.

Figure~\ref{fig:agentic_loop} illustrates an infinite loop extracted from task \texttt{django\_\_django-15525} of SWE-bench Verified.
When the L31-8B model encounters a reasoning bottleneck, it enters a repetitive action pattern, repeatedly generating identical Thought and Action sequences.
Unlike simple syntactic repetition, this represents a higher-level Agentic Action Loop.
Suffix Decoding, relying on token-level redundancy, identifies the recurring reasoning trace as a high-confidence pattern, thereby achieving a near-perfect match rate by retrieving historical sequences. In contrast, L31-70B generates more divergent tokens in this example, which reduces the match with the preceding context and leads to a lower acceptance rate.

The impact of infinite loops extends beyond model-free methods and can also affect model-based approaches.
On SWE-bench, Eagle-3 shows unusually low speculation efficiency on Qwen3-30A3 and L31-8B, where the mean acceptance length drops to $\tau=0.43$ and $\tau=0.64$, respectively, compared with $\tau=1.87$ on Qwen3-32B and $\tau=1.07$ on L31-70B.
These low values suggest that degenerate or highly repetitive agentic traces can disrupt draft-target alignment for model-based SD as well.

The occurrence of ``Infinite Loops'' represents a widespread phenomenon rather than an isolated anomaly in current LLM-based agents.
To investigate this further, we conducted a quantitative analysis of the execution trajectories within the SWE-bench Verified dataset.
We define an infinite loop as a trajectory in which the agent produces the same output block at least five times while solving a single instance.
As illustrated in Figure~\ref{fig:num_of_loops}, the L31-8B model encounters infinite loops in 226 distinct tasks.
In a similar vein, the Qwen3-32B model demonstrates this pattern in 210 tasks.
Qwen3-30A3 also exhibits this behavior in 167 tasks.
Conversely, L31-70B exhibits a notably lower frequency, with only 89 tasks affected by such repetitive generations.
These quantitative findings suggest that repetitive agentic behavior varies substantially across target models and can become a major confounding factor in SD evaluation.

This ubiquity necessitates a shift in how we evaluate SD in agentic applications.
It suggests that infinite loops are not merely isolated anomalies, but a recurring failure mode in agentic SE evaluation.
Consequently, relying solely on intrinsic acceleration metrics is insufficient and potentially misleading, as the higher predictability of these loops can introduce noise into speedup measurements without reflecting meaningful task progress.

\begin{findingbox}
\textbf{Finding 6:} The prevalence of ``infinite loops'' in complex agentic scenarios introduces noise into the evaluation of SD.
\end{findingbox}

\subsection{RQ3: Comparison of SD between Natural Language and Code}
\label{subsec:rq3}

\begin{table}[h]
    \centering
    \caption{Experimental results on MT-bench compared with SE tasks. The SE column reports the arithmetic mean of performance metrics across the six evaluated SE benchmarks.}

\label{tab:mt_bench_results}
    \begin{tabular}{llcccc}
        \toprule
        \multirow{2}{*}{\textbf{Model}} & \multirow{2}{*}{\textbf{Method}} & \multicolumn{2}{c}{\textbf{MT-bench}} & \multicolumn{2}{c}{\textbf{SE}} \\
        \cmidrule(lr){3-4} \cmidrule(lr){5-6}
         & & \textbf{Speedup} & $\tau$ & \textbf{Speedup} & $\tau$ \\
        \midrule
        
        \multirow{3}{*}{Qwen3-32B} 
          & PLD & 0.94$\times$ & 1.70 & 1.42$\times$ & 3.23 \\
          & Suffix & 1.05$\times$ & 0.44 & 1.90$\times$ & 2.43 \\
          & Eagle-3 & 1.69$\times$ & 1.56 & 1.57$\times$ & 2.00 \\
        \midrule

        \multirow{3}{*}{Qwen3-30A3} 
          & PLD & 0.85$\times$ & 1.82 & 1.15$\times$ & 3.09 \\
          & Suffix & 0.74$\times$ & 0.40 & 1.34$\times$ & 2.37 \\
          & Eagle-3 & 0.66$\times$ & 1.27 & 0.72$\times$ & 1.57 \\
        \midrule

        \multirow{3}{*}{L31-8B} 
          & PLD & 1.09$\times$ & 2.27 & 1.51$\times$ & 3.13 \\
          & Suffix & 1.27$\times$ & 0.64 & 2.34$\times$ & 2.77 \\
          & Eagle-3 & 1.95$\times$ & 1.74 & 1.61$\times$ & 1.63 \\
        \midrule

        \multirow{3}{*}{L31-70B} 
          & PLD & 0.96$\times$ & 2.19 & 1.25$\times$ & 3.12 \\
          & Suffix & 1.05$\times$ & 0.60 & 1.56$\times$ & 2.61 \\
          & Eagle-3 & 0.97$\times$ & 0.39 & 1.16$\times$ & 1.00 \\

        \bottomrule
    \end{tabular}%

\end{table}
\phead{Method Robustness on Natural Language.}
To answer RQ3, we employed the same experimental setup on MT-bench~\cite{zheng2023judgingllmasajudgemtbenchchatbot}, a representative natural language benchmark. MT-bench is a widely used benchmark designed to evaluate the conversational and instruction-following capabilities of LLMs. It comprises high-quality multi-turn questions across diverse categories, where the response quality is typically scored by a judge model. The SE results in Table~\ref{tab:mt_bench_results} are averaged across the six evaluated SE benchmarks.

The results show that acceptance lengths drop markedly on MT-bench for model-free methods than for Eagle-3. Across the four models, PLD's average $\tau$ decreases from 3.14 on SE tasks to 2.00 on MT-bench, while Suffix Decoding decreases more sharply from 2.55 to 0.52. Such short acceptance lengths make it difficult for model-free methods to translate drafted tokens into tangible acceleration. By contrast, Eagle-3 exhibits a smaller decrease in $\tau$, suggesting stronger robustness on natural language tasks.

\begin{figure*}[t]
    \centering
    \begin{subfigure}[b]{0.48\textwidth}
        \centering
        \includegraphics[width=\linewidth]{figures/n-alpha/LiveCodeBench.pdf}
        \caption{\textbf{LiveCodeBench}}
        \label{fig:n_alpha_lcb_mt}
    \end{subfigure}
    \hfill 
    \begin{subfigure}[b]{0.48\textwidth}
        \centering
        \includegraphics[width=\linewidth]{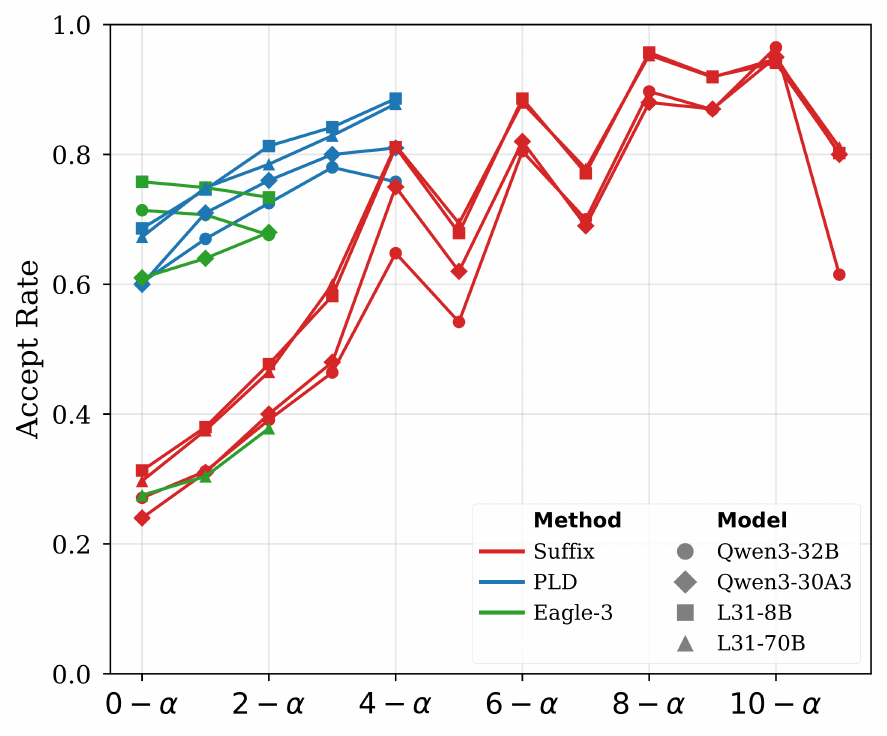}
        \caption{\textbf{MT-bench}}
        \label{fig:n_alpha_mt}
    \end{subfigure}
    
    \caption{The $n-\alpha$ curves on (a) LiveCodeBench and (b) MT-bench. }
    \label{fig:n_alpha_comparison}
\end{figure*}

To better understand why model-free methods degrade on MT-bench, we analyze the $n-\alpha$ curves in Figure~\ref{fig:n_alpha_comparison} and Figure~\ref{fig:n_alpha_all}.
Interestingly, both PLD and Suffix Decoding exhibit upward trends on MT-bench, similar to their behavior on SE tasks.
This trajectory is determined by their draft generation mechanism, which relies on retrieving tokens from context history.
However, compared with SE tasks, MT-bench appears to provide fewer long contextual spans that can be reused by context-retrieval-based drafting.
Consequently, even though the acceptance rate increases when a long match is found, such matches occur less frequently in natural language.
This reduces the overall $\tau$ and weakens the effectiveness of model-free methods.

This contrast is consistent with the mechanism of Eagle-3. Unlike model-free methods, Eagle-3 does not rely on exact contextual repetitions; instead, its drafter approximates the target distribution. Excluding L31-70B, whose Eagle-3 drafter is trained separately in our study, Eagle-3 maintains comparable acceptance lengths between SE tasks and MT-bench: its average $\tau$ changes from 1.73 on SE tasks to 1.52 on MT-bench across Qwen3-32B, Qwen3-30A3, and L31-8B. This suggests that model-based methods are less sensitive to the loss of long repeated spans in natural language than model-free methods.

\begin{findingbox}
\textbf{Finding 7:} Model-based SD methods are more robust than model-free methods on natural language tasks.
\end{findingbox}

\phead{Quantitative Analysis of Contextual Redundancy.}
To empirically illustrate the stronger long-range repetitiveness of code tasks, we extended our analysis to the entire datasets of LiveCodeBench and MT-bench. We define the output repetition ratio as the proportion of output-side $N$-grams that appear more than once in the full context, including both the prompt prefix and the generated output. The dataset-level value is the macro-average of this ratio across all samples.

\begin{table}[h]
    \vspace{-1em}
    \centering
    \caption{Comparison of local repetitiveness between LiveCodeBench and MT-bench for $N \in \{2, 3, 4, 5\}$. The metric represents the percentage of $N$-grams that are part of a repeating sequence within the context.}
    \label{tab:repetition}
        \begin{tabular}{lcccc}
        \toprule
        \multirow{2}{*}{\textbf{Domain (Task)}} & \multicolumn{4}{c}{\textbf{Number of repetitions (\%)}} \\
        \cmidrule(lr){2-5}
         & \textbf{N=2} & \textbf{N=3} & \textbf{N=4} & \textbf{N=5} \\
        \midrule
        \textbf{Code} & 28.64\% & 17.66\% & 12.68\% & 9.62\% \\
        \textbf{Natural Language} & 28.66\% & 15.40\% & 9.63\% & 6.85\% \\
        \bottomrule
        \end{tabular}
    \vspace{-0.5em}
\end{table}
As shown in Table~\ref{tab:repetition}, the two tasks are similar at $N=2$, but code retains more long-range repetitions as $N$ increases. At $N=3$, code reaches 17.66\%, compared with 15.40\% for natural language. This gap widens further at $N=4$ and $N=5$, indicating that code provides more reusable contextual spans for model-free methods.

Furthermore, SE tasks contain substantial text alongside code. For instance, code tokens and text tokens account for 35.17\% and 64.83\% of the SWE-bench dataset, respectively.\footnote{These token proportions are computed from the Qwen3-32B SWE-bench execution traces.}
Despite this high proportion of text, model-free methods still achieve higher mean acceptance length on SWE-bench than on MT-bench. This suggests that the textual portions in SE tasks are more structured and reusable than natural language.

\begin{findingbox}
\textbf{Finding 8:} Compared with MT-bench, SE tasks provide more reusable contextual patterns, which helps model-free SD achieve stronger acceleration.
\end{findingbox}

\phead{Capability to Capture Repetitions via Domain-Specific Training.}
We observe an interesting phenomenon regarding Eagle-3 on L31-70B-Instruct.
Unlike the downward trends exhibited by the other RedHatAI draft models, the custom Eagle-3 draft model for L31-70B displays an ascending $n-\alpha$ curve on both MT-bench and SE tasks (shown in Figure~\ref{fig:n_alpha_all} and Figure~\ref{fig:n_alpha_comparison}), mirroring the behavior of model-free methods.
We attribute this distinct behavioral divergence to the specific composition of the training data. The Eagle-3 drafter for the 70B model was trained on SE datasets (BAAI/TACO and SWE-Gym/SWE-Gym), whereas the drafters for the smaller models were trained on general natural language corpora (ShareGPT and UltraChat200k).
This suggests that training on SE data may help the model capture repetition-oriented patterns.

\begin{findingbox}
\textbf{Finding 9:} Model-based methods trained on SE datasets exhibit stronger repetition-oriented drafting behavior.
\end{findingbox}

\section{Discussion}
\label{sec:discussion}

\subsection{Parameter Sensitivity of SD}
\label{subsec:para_of_SD}
We further investigate the sensitivity of inference acceleration to the number of draft tokens $K$. 
Given that SE tasks exhibit stronger contextual redundancy compared to natural language, the optimal parameters for SD in SE scenarios should be more aggressive than those in natural language.
Based on this, we conducted an experiment on Eagle-3 with Qwen3-32B by increasing the number of speculative tokens from the default value of 3 to 5.

\begin{table*}[h]
    \centering
    \caption{Performance sensitivity of Eagle-3 (Qwen3-32B) to number of speculative tokens ($K$). 
    Comparison between default ($K=3$) and aggressive ($K=5$) settings. 
    \textbf{Bold} indicates the best performance.}
    \label{tab:eagle_k5}
    \resizebox{1\textwidth}{!}{%
    \begin{tabular}{lcccccccccccccc}
        \toprule
        \multirow{2}{*}{\textbf{Configuration}} & \multicolumn{2}{c}{\textbf{LiveCodeBench}} & \multicolumn{2}{c}{\textbf{HumanEval}} & \multicolumn{2}{c}{\textbf{SWE-bench}} & \multicolumn{2}{c}{\textbf{Defects4J}} & \multicolumn{2}{c}{\textbf{Aider Polyglot}} & \multicolumn{2}{c}{\textbf{CanItEdit}} & \multicolumn{2}{c}{\textbf{MT-bench}} \\
        \cmidrule(lr){2-3} \cmidrule(lr){4-5} \cmidrule(lr){6-7} \cmidrule(lr){8-9} \cmidrule(lr){10-11} \cmidrule(lr){12-13} \cmidrule(lr){14-15}
         & \textbf{Speedup} & $\tau$ & \textbf{Speedup} & $\tau$ & \textbf{Speedup} & $\tau$ & \textbf{Speedup} & $\tau$ & \textbf{Speedup} & $\tau$ & \textbf{Speedup} & $\tau$ & \textbf{Speedup} & $\tau$ \\
        \midrule
        $K=3$  & 1.71$\times$ & 1.87 & 1.94$\times$ & 1.99 & 1.10$\times$ & 1.87 & 1.21$\times$ & 2.10 & 1.29$\times$ & 1.81 & 2.14$\times$ & 2.35 & 1.69$\times$ & 1.56 \\
        $K=5$  & \textbf{1.79$\times$} & \textbf{2.25} & \textbf{2.26$\times$} & \textbf{2.55} & \textbf{1.53$\times$} & \textbf{2.39} & \textbf{2.23$\times$} & \textbf{2.83} & \textbf{1.43$\times$} & \textbf{2.27} & \textbf{2.56$\times$} & \textbf{3.32} & \textbf{1.76$\times$} & \textbf{1.86} \\
        \bottomrule
    \end{tabular}%
    }
\end{table*}

As presented in Table~\ref{tab:eagle_k5}, the aggressive configuration ($K=5$) yields consistent performance gains across both SE and natural language tasks. However, the improvements are more pronounced in the SE domain. Across the six SE benchmarks, the mean speedup and mean acceptance length ($\tau$) increase by approximately 25.7\% (from 1.57$\times$ to 1.97$\times$) and 30.2\% (from 2.00 to 2.60), respectively. In contrast, on the natural language task, the gains are relatively modest, with speedup and $\tau$ increasing by 4.1\% (from 1.69$\times$ to 1.76$\times$) and 19.2\% (from 1.56 to 1.86), respectively. Notably, on the repository-level SWE-bench task, the speedup increases substantially from 1.10$\times$ to 1.53$\times$, and similar gains appear on Defects4J and CanItEdit.

This observation indicates that due to the higher predictability and contextual redundancy of SE tasks compared to natural language, SD in SE tasks can accommodate more aggressive hyperparameter settings than the standard configurations used in natural language.

\subsection{Implications}
\label{subsec:implications}

Based on the findings from our research questions, we discuss the following implications for the SE community:

\phead{The Strategy of Deployment and Model Alignment.} 
Our results in RQ1 and RQ3 reveal that the effectiveness of SD is highly dependent on the choice of decoding strategy and the alignment of the draft model. 

Our experiments indicate that model-free methods often provide practical acceleration in SE tasks, particularly when the workload contains reusable contextual redundancy, such as in repository-level editing and repair. However, their benefits are not universal across all model-task combinations. Due to their simple deployment, zero training cost, and absence of additional GPU memory overhead for draft inference, we still recommend prioritizing these methods for resource-constrained environments, such as local IDE assistants.

In contrast, model-based methods can be effective in fresh code generation tasks where context history is limited, but their benefits depend on draft-target and task alignment.
Furthermore, our discussion in Section \S\ref{subsec:para_of_SD} indicates that SE tasks can accommodate more aggressive hyperparameters than standard natural language tasks due to the higher predictability. For example, increasing the number of draft tokens yields significantly higher marginal gains in SE tasks.
However, we observed that drafters trained on general chat corpora may suffer from distribution shifts when applied to SE tasks, leading to performance degradation or even negative speedups. Therefore, practitioners who adopt model-based SD for SE workloads should carefully validate draft-target alignment and, when necessary, consider domain-specific training or adaptation to bridge this distributional gap.

\phead{The Evaluation of SD in Agentic Tasks.}
In RQ2, we identified the prevalence of ``infinite loops'' in agentic workflows, where agents encountering reasoning deadlocks generate identical thought and action blocks repeatedly. This phenomenon introduces noise into SD evaluation, as such high context redundancy can favor model-free methods that rely on retrieving contextual repetitions. To further examine this issue, we compare task-specific performance scores with the mean acceptance length of different SD methods in Table~\ref{tab:correlation_data}.

\begin{table*}[h]
    \centering
    \caption{Comparison between task performance and SD acceptance length. Performance Score denotes the task-specific metric, while Mean Acceptance Length ($\tau$) indicates the average number of accepted tokens per step.}
    \label{tab:correlation_data}
    \setlength{\tabcolsep}{12pt}
    \resizebox{\textwidth}{!}{%
    \begin{tabular}{llcccc}
        \toprule
        \multirow{2}{*}{\textbf{Benchmark}} & \multirow{2}{*}{\textbf{Model}} & \multirow{2}{*}{\textbf{Performance Score}} & \multicolumn{3}{c}{\textbf{Mean Accept Length ($\tau$)}} \\
        \cmidrule(lr){4-6}\textit{Metrics}& & & \textbf{PLD} & \textbf{Suffix} & \textbf{Eagle-3} \\
        \midrule
        
        \multirow{4}{*}{\shortstack[l]{\textbf{LiveCodeBench}\\\textit{ Pass@1} (\%)}} & Qwen3-32B     & 46.1 & 2.89 & 1.21 & 1.87 \\
          & Qwen3-30A3     & 56.0 & 2.21 & 1.02 & 1.75 \\
          & L31-8B   & 17.0 & 2.36 & 1.30 & 1.82 \\
          & L31-70B  & 30.5 & 2.39 & 1.08 & 1.74 \\
        \midrule

        \multirow{4}{*}{\shortstack[l]{\textbf{HumanEval}\\\textit{Pass@1}
        (\%)}} 
          & Qwen3-32B     & 83.5 & 2.89 & 1.10 & 1.99 \\
          & Qwen3-30A3    & 96.3 & 2.57 & 0.91 & 1.93 \\
          & L31-8B   & 68.9 & 2.81 & 1.13 & 2.11 \\
          & L31-70B  & 72.6 & 2.40 & 1.00 & 1.27 \\
        \midrule

        \multirow{4}{*}{\shortstack[l]{\textbf{SWE-bench}\\\textit{Resolved} (\%)}} 
          & Qwen3-32B     & 12.4 & 2.92 & 3.48 & 1.87 \\
          & Qwen3-30A3     & 10.4 & 3.25 & 4.78 & 0.43 \\
          & L31-8B   & 1.4 & 2.99 & 5.48 & 0.64 \\
          & L31-70B  & 7.2 & 2.69 & 3.39 & 1.07 \\
        \midrule

        \multirow{4}{*}{\shortstack[l]{\textbf{Defects4J}\\\textit{Correct} (\%)}} 
          & Qwen3-32B     & 18.54 & 3.40 & 3.33 & 2.10 \\
          & Qwen3-30A3    & 18.08 & 3.38 & 3.14 & 1.57 \\
          & L31-8B   & 7.09 & 3.23 & 3.18 & 1.56 \\
          & L31-70B  & 19.68 & 3.77 & 3.55 & 0.35 \\
        \midrule

        \multirow{4}{*}{\shortstack[l]{\textbf{Aider Polyglot}\\\textit{Pass Rate 1 \& 2} (\%)}} 
          & Qwen3-32B     & 18.2 \& 41.3 & 3.19 & 1.35 & 1.81 \\
          & Qwen3-30A3     & 11.6 \& 25.8 & 3.29 & 1.41 & 1.52 \\
          & L31-8B   & 1.3 \& 2.7 & 3.36 & 1.99 & 1.49 \\
          & L31-70B  & 1.8 \& 7.1 & 3.40 & 2.72 & 0.48 \\
        \midrule

        \multirow{4}{*}{\shortstack[l]{\textbf{CanItEdit}\\\textit{Pass@1}
        (\%)}} 
          & Qwen3-32B     & 58.1 & 4.07 & 4.11 & 2.35 \\
          & Qwen3-30A3    & 54.8 & 3.83 & 2.93 & 2.22 \\
          & L31-8B   & 43.3 & 4.01 & 3.54 & 2.18 \\
          & L31-70B  & 53.3 & 4.05 & 3.93 & 1.09 \\
          
        \bottomrule
    \end{tabular}%
    }
\end{table*}

%
Table~\ref{tab:correlation_data} suggests that, in agentic tasks, lower task performance can coincide with higher acceptance lengths for model-free methods. This pattern is particularly clear on Aider Polyglot: Qwen3-32B achieves the best task score (18.2\% and 41.3\%) but has lower PLD and Suffix acceptance lengths ($\tau=3.19$ and $\tau=1.35$), while L31-70B obtains much lower task scores (1.8\% and 7.1\%) but higher PLD and Suffix acceptance lengths ($\tau=3.40$ and $\tau=2.72$). On SWE-bench, L31-8B has the lowest problem-solving capability (1.4\% Resolved) but reaches the highest Suffix Decoding acceptance length ($\tau=5.48$). Similarly, Qwen3-30A3 achieves a lower task score than Qwen3-32B (10.4\% vs. 12.4\%) but obtains a higher PLD acceptance length ($\tau=3.25$ vs. 2.92). These examples suggest that model-free acceptance statistics can be inflated by repetitive or degenerate agentic traces, rather than reflecting meaningful task progress alone.

Eagle-3 does not exhibit the same pattern consistently, because its acceptance length is more affected by draft-target alignment than by direct contextual reuse. For example, on SWE-bench, Qwen3-32B obtains $\tau=1.87$, while Qwen3-30A3 and L31-8B obtain much lower values ($\tau=0.43$ and $\tau=0.64$). On Aider Polyglot, Qwen3-32B and Qwen3-30A3 also achieve higher Eagle-3 acceptance lengths than L31-70B. Therefore, when evaluating SD on complex agentic tasks, researchers should interpret acceleration metrics together with task-specific quality metrics and method-specific drafting mechanisms.

\phead{Verification Efficiency in Long-Context Regimes.}
Our investigation into repository-level benchmarks highlights a phenomenon of diminishing returns where high acceptance rates (implied by $\tau$) do not linearly translate into speedup gains.
In tasks like SWE-bench where context windows often exceed 16k tokens, the verification process transitions into a compute-bound regime, which is driven by the significant overhead of attention mechanisms and KV cache management.
Consequently, even accurate draft models may fail to deliver effective acceleration if the verification latency remains high.
This implies that solely optimizing the acceptance rate of the draft model is insufficient for repository-level SE assistants.
Future research and system design should place more emphasis on optimizing verification efficiency.

\phead{The practical impact of SD on SE.}
In SE scenarios, the Cursor editor\footnote{\url{https://cursor.sh/blog/1000-tokens}} adopts a highly optimized approach for SD. Specifically, rather than relying on an auxiliary draft model, it leverages the local user context as a deterministic prior for draft generation, which are subsequently verified by a customized target model to achieve rapid code edits. This integration effectively exploits contextual redundancy while maintaining high generation quality. While the practical deployment of SD still faces challenges regarding computational resource limits and long context processing, this technology demonstrates clear potential. It can accelerate inference and transition intelligent programming tools from conventional dialogue-based systems to low-latency code completion assistants. This reduction in latency directly enhances the responsiveness of programming tools and makes it feasible to deploy LLMs in agent workflows that require rapid and iterative interactions.

\subsection{Threats to Validity}
\label{subsec:threats}

We identify the following threats to validity of our study:

\begin{itemize}
    \item \textbf{Limited Benchmarks and Models.} 
    The empirical experiments presented in this study are conducted on a selection of six specific benchmarks and two distinct model families.
    We acknowledge that this constrained scope may potentially limit the broader generalizability of our derived findings to other untested scenarios.
    To mitigate this, we meticulously selected widely adopted and representative benchmarks that cover the primary scenarios in SE (i.e., code generation, code repair, and code editing).
    Furthermore, we employed two influential model families (Llama-3.1 and Qwen3) spanning diverse parameter scales and architectures, including both dense and sparsely activated MoE models, ensuring that our conclusions remain robust across varying model architectures and capacities.

    \item \textbf{Mismatch in Training Data and Configurations.}
    The Eagle-3 draft models in our evaluation are primarily trained on natural language datasets and used with default hyperparameters, which might fail to represent the method's optimal performance in the SE domain.
    To address this, we extended our analysis by training a specialized draft model using SE-specific datasets and conducting sensitivity analyses with varying hyperparameters.
    
    \item \textbf{Generalizability of Decoding Strategies.}
    We primarily utilized greedy decoding ($T=0$), whereas real-world coding assistants often employ stochastic sampling ($T>0$) to encourage diversity.
    Theoretical analysis suggests that as sampling temperature increases, the divergence between the draft and target output distributions widens, inherently reducing the acceptance rate of speculative tokens~\cite{leviathan2023fast}. Consequently, our results under greedy settings serve as a rigorous \textit{upper-bound baseline}, providing a theoretical performance ceiling for system designers.
\end{itemize}

\section{Conclusion}
\label{sec:conclusion}

In this paper, we present a systematic empirical study to explore the effectiveness and mechanisms of SD across six benchmarks covering three representative SE task categories.
Our experiments are conducted across a diverse spectrum of SD strategies, and the results demonstrate that this paradigm serves as an effective approach to reduce inference latency, albeit with notable variance across different experimental settings. Additionally, we observe that models deployed in complex agentic scenarios could fall into infinite loops, a phenomenon that introduces noise into the evaluation of SD.
Compared with natural language, SE tasks provide more useful long-range contextual repetitions, enabling simple, training-free methods to achieve practical acceleration in suitable scenarios.
We further discuss the impact of hyperparameters and provide guidelines based on our findings, expecting future research in inference acceleration for SE tasks to benefit from our insights.

\section*{Data Availability}
To support open science, we make the replication package with source code publicly available at: \url{https://github.com/AltriaSetsuna/speculative-decoding-empirical}.

\begin{acks}
This work was supported by the National Key R\&D Program of China (No. 2024YFB4506400) and CCF-Huawei Populus Grove Fund.
\end{acks}

\bibliographystyle{ACM-Reference-Format}
\bibliography{references}

@misc{qwen3technicalreport,
          title={Qwen3 Technical Report}, 
          author={Qwen Team},
          year={2025},
          eprint={2505.09388},
          archivePrefix={arXiv},
          primaryClass={cs.CL},
          url={https://arxiv.org/abs/2505.09388}, 
    }

@misc{deepseek2024v2,
          title={DeepSeek-Coder-V2: Breaking the Barrier of Closed-Source Models in Code Intelligence}, 
          author={DeepSeek-AI and Qihao Zhu and Daya Guo and Zhihong Shao and Dejian Yang and Peiyi Wang and Runxin Xu and Y. Wu and Yukun Li and Huazuo Gao and Shirong Ma and Wangding Zeng and Xiao Bi and Zihui Gu and Hanwei Xu and Damai Dai and Kai Dong and Liyue Zhang and Yishi Piao and Zhibin Gou and Zhenda Xie and Zhewen Hao and Bingxuan Wang and Junxiao Song and Deli Chen and Xin Xie and Kang Guan and Yuxiang You and Aixin Liu and Qiushi Du and Wenjun Gao and Xuan Lu and Qinyu Chen and Yaohui Wang and Chengqi Deng and Jiashi Li and Chenggang Zhao and Chong Ruan and Fuli Luo and Wenfeng Liang},
          year={2024},
          eprint={2406.11931},
          archivePrefix={arXiv},
          primaryClass={cs.SE},
          url={https://arxiv.org/abs/2406.11931}, 
    }

@misc{lozhkov2024starcoder2,
          title={StarCoder 2 and The Stack v2: The Next Generation}, 
          author={Anton Lozhkov and Raymond Li and Loubna Ben Allal and Federico Cassano and Joel Lamy-Poirier and Nouamane Tazi and Ao Tang and Dmytro Pykhtar and Jiawei Liu and Yuxiang Wei and Tianyang Liu and Max Tian and Denis Kocetkov and Arthur Zucker and Younes Belkada and Zijian Wang and Qian Liu and Dmitry Abulkhanov and Indraneil Paul and Zhuang Li and Wen-Ding Li and Megan Risdal and Jia Li and Jian Zhu and Terry Yue Zhuo and Evgenii Zheltonozhskii and Nii Osae Osae Dade and Wenhao Yu and Lucas Krauß and Naman Jain and Yixuan Su and Xuanli He and Manan Dey and Edoardo Abati and Yekun Chai and Niklas Muennighoff and Xiangru Tang and Muhtasham Oblokulov and Christopher Akiki and Marc Marone and Chenghao Mou and Mayank Mishra and Alex Gu and Binyuan Hui and Tri Dao and Armel Zebaze and Olivier Dehaene and Nicolas Patry and Canwen Xu and Julian McAuley and Han Hu and Torsten Scholak and Sebastien Paquet and Jennifer Robinson and Carolyn Jane Anderson and Nicolas Chapados and Mostofa Patwary and Nima Tajbakhsh and Yacine Jernite and Carlos Muñoz Ferrandis and Lingming Zhang and Sean Hughes and Thomas Wolf and Arjun Guha and Leandro von Werra and Harm de Vries},
          year={2024},
          eprint={2402.19173},
          archivePrefix={arXiv},
          primaryClass={cs.SE},
          url={https://arxiv.org/abs/2402.19173}, 
    }

@article{grattafiori2024llama,
          title={The llama 3 herd of models},
          author={Grattafiori, Aaron and Dubey, Abhimanyu and Jauhri, Abhinav and Pandey, Abhinav and Kadian, Abhishek and Al-Dahle, Ahmad and Letman, Aiesha and Mathur, Akhil and Schelten, Alan and Vaughan, Alex and others},
          journal={arXiv preprint arXiv:2407.21783},
          year={2024}
        }

@article{yang2024swe,
          title={Swe-agent: Agent-computer interfaces enable automated software engineering},
          author={Yang, John and Jimenez, Carlos E and Wettig, Alexander and Lieret, Kilian and Yao, Shunyu and Narasimhan, Karthik and Press, Ofir},
          journal={Advances in Neural Information Processing Systems},
          volume={37},
          pages={50528--50652},
          year={2024}
}

@inproceedings{wang2024openhands,
  title={OpenHands: An Open Platform for AI Software Developers as Generalist Agents},
  author={Wang, Xingyao and Li, Boxuan and Song, Yufan and Xu, Frank F and Tang, Xiangru and Zhuge, Mingchen and Pan, Jiayi and Song, Yueqi and Li, Bowen and Singh, Jaskirat and others},
  booktitle={The Thirteenth International Conference on Learning Representations}
}

@article{hou2024large,
          title={Large language models for software engineering: A systematic literature review},
          author={Hou, Xinyi and Zhao, Yanjie and Liu, Yue and Yang, Zhou and Wang, Kailong and Li, Li and Luo, Xiapu and Lo, David and Grundy, John and Wang, Haoyu},
          journal={ACM Transactions on Software Engineering and Methodology},
          volume={33},
          number={8},
          pages={1--79},
          year={2024},
          publisher={ACM New York, NY},
          doi={10.1145/3695988}
        }

@inproceedings{frantar2022gptq,
  title={OPTQ: Accurate Quantization for Generative Pre-trained Transformers},
  author={Frantar, Elias and Ashkboos, Saleh and Hoefler, Torsten and Alistarh, Dan},
  booktitle={The Eleventh International Conference on Learning Representations}
}

@inproceedings{dettmers2022gpt3,
 author = {Dettmers, Tim and Lewis, Mike and Belkada, Younes and Zettlemoyer, Luke},
 booktitle = {Advances in Neural Information Processing Systems},
 doi = {10.52202/068431-2198},
 editor = {S. Koyejo and S. Mohamed and A. Agarwal and D. Belgrave and K. Cho and A. Oh},
 pages = {30318--30332},
 publisher = {Curran Associates, Inc.},
 title = {GPT3.int8(): 8-bit Matrix Multiplication for Transformers at Scale},
 url = {https://proceedings.neurips.cc/paper_files/paper/2022/file/c3ba4962c05c49636d4c6206a97e9c8a-Paper-Conference.pdf},
 volume = {35},
 year = {2022}
}

@article{lin2024awq,
          title={Awq: Activation-aware weight quantization for on-device llm compression and acceleration},
          author={Lin, Ji and Tang, Jiaming and Tang, Haotian and Yang, Shang and Chen, Wei-Ming and Wang, Wei-Chen and Xiao, Guangxuan and Dang, Xingyu and Gan, Chuang and Han, Song},
          journal={Proceedings of machine learning and systems},
          volume={6},
          pages={87--100},
          year={2024},
          doi = {10.1145/3714983.3714987},
        }

@inproceedings{frantar2023sparsegpt,
          title={Sparsegpt: Massive language models can be accurately pruned in one-shot},
          author={Frantar, Elias and Alistarh, Dan},
          booktitle={International conference on machine learning},
          pages={10323--10337},
          year={2023},
          organization={PMLR}
        }

@inproceedings{leviathan2023fast,
          title={Fast inference from transformers via speculative decoding},
          author={Leviathan, Yaniv and Kalman, Matan and Matias, Yossi},
          booktitle={International Conference on Machine Learning},
          pages={19274--19286},
          year={2023},
          organization={PMLR}
        }

@article{chen2023accelerating,
          title={Accelerating large language model decoding with speculative sampling},
          author={Chen, Charlie and Borgeaud, Sebastian and Irving, Geoffrey and Lespiau, Jean-Baptiste and Sifre, Laurent and Jumper, John},
          journal={arXiv preprint arXiv:2302.01318},
          year={2023}
        }

@misc{ankner2024hydrasequentiallydependentdraftheads,
          title={Hydra: Sequentially-Dependent Draft Heads for Medusa Decoding}, 
          author={Zachary Ankner and Rishab Parthasarathy and Aniruddha Nrusimha and Christopher Rinard and Jonathan Ragan-Kelley and William Brandon},
          year={2024},
          eprint={2402.05109},
          archivePrefix={arXiv},
          primaryClass={cs.LG},
          url={https://arxiv.org/abs/2402.05109}, 
    }

@inproceedings{he2024restretrievalbasedspeculativedecoding,
  title={REST: Retrieval-Based Speculative Decoding},
  author={He, Zhenyu and Zhong, Zexuan and Cai, Tianle and Lee, Jason D and He, Di},
  booktitle={2024 Conference of the North American Chapter of the Association for Computational Linguistics: Human Language Technologies, NAACL 2024},
  pages={1582--1595},
  year={2024},
  organization={Association for Computational Linguistics (ACL)},
  doi = "10.18653/v1/2024.naacl-long.88",
}

@inproceedings{fu2024breaksequentialdependencyllm,
  title={Break the Sequential Dependency of LLM Inference Using Lookahead Decoding},
  author={Fu, Yichao and Bailis, Peter and Stoica, Ion and Zhang, Hao},
  booktitle={Forty-first International Conference on Machine Learning}
}

@inproceedings{cai2024medusasimplellminference,
  title={Medusa: Simple LLM Inference Acceleration Framework with Multiple Decoding Heads},
  author={Cai, Tianle and Li, Yuhong and Geng, Zhengyang and Peng, Hongwu and Lee, Jason D and Chen, Deming and Dao, Tri},
  booktitle={Forty-first International Conference on Machine Learning}
}

@misc{wertheimer2024acceleratingproductionllmscombined,
          title={Accelerating Production LLMs with Combined Token/Embedding Speculators}, 
          author={Davis Wertheimer and Joshua Rosenkranz and Thomas Parnell and Sahil Suneja and Pavithra Ranganathan and Raghu Ganti and Mudhakar Srivatsa},
          year={2024},
          eprint={2404.19124},
          archivePrefix={arXiv},
          primaryClass={cs.CL},
          url={https://arxiv.org/abs/2404.19124}, 
    }

@inproceedings{li2025eaglespeculativesamplingrequires,
  title={EAGLE: Speculative Sampling Requires Rethinking Feature Uncertainty},
  author={Li, Yuhui and Wei, Fangyun and Zhang, Chao and Zhang, Hongyang},
  booktitle={Forty-first International Conference on Machine Learning}
}

@misc{li2024eagle2fasterinferencelanguage,
          title={EAGLE-2: Faster Inference of Language Models with Dynamic Draft Trees}, 
          author={Yuhui Li and Fangyun Wei and Chao Zhang and Hongyang Zhang},
          year={2024},
          eprint={2406.16858},
          archivePrefix={arXiv},
          primaryClass={cs.CL},
          url={https://arxiv.org/abs/2406.16858}, 
    }

@misc{li2025eagle3scalinginferenceacceleration,
          title={EAGLE-3: Scaling up Inference Acceleration of Large Language Models via Training-Time Test}, 
          author={Yuhui Li and Fangyun Wei and Chao Zhang and Hongyang Zhang},
          year={2025},
          eprint={2503.01840},
          archivePrefix={arXiv},
          primaryClass={cs.CL},
          url={https://arxiv.org/abs/2503.01840}, 
    }

@misc{saxena2023prompt,
        title = {Prompt Lookup Decoding},
        author = {Apoorv Saxena},
        year = {2023},
        month = {November},
        howpublished = {\url{https://github.com/apoorvumang/prompt-lookup-decoding/}}
    }

@inproceedings{oliaro2025suffixdecodingextremespeculativedecoding,
  title={Suffixdecoding: Extreme speculative decoding for emerging ai applications},
  author={Oliaro, Gabriele and Jia, Zhihao and Campos, Daniel F and Qiao, Aurick},
  booktitle={The Thirty-ninth Annual Conference on Neural Information Processing Systems}
}

@misc{chen2021evaluatinglargelanguagemodels,
          title={Evaluating Large Language Models Trained on Code}, 
          author={Mark Chen and Jerry Tworek and Heewoo Jun and Qiming Yuan and Henrique Ponde de Oliveira Pinto and Jared Kaplan and Harri Edwards and Yuri Burda and Nicholas Joseph and Greg Brockman and Alex Ray and Raul Puri and Gretchen Krueger and Michael Petrov and Heidy Khlaaf and Girish Sastry and Pamela Mishkin and Brooke Chan and Scott Gray and Nick Ryder and Mikhail Pavlov and Alethea Power and Lukasz Kaiser and Mohammad Bavarian and Clemens Winter and Philippe Tillet and Felipe Petroski Such and Dave Cummings and Matthias Plappert and Fotios Chantzis and Elizabeth Barnes and Ariel Herbert-Voss and William Hebgen Guss and Alex Nichol and Alex Paino and Nikolas Tezak and Jie Tang and Igor Babuschkin and Suchir Balaji and Shantanu Jain and William Saunders and Christopher Hesse and Andrew N. Carr and Jan Leike and Josh Achiam and Vedant Misra and Evan Morikawa and Alec Radford and Matthew Knight and Miles Brundage and Mira Murati and Katie Mayer and Peter Welinder and Bob McGrew and Dario Amodei and Sam McCandlish and Ilya Sutskever and Wojciech Zaremba},
          year={2021},
          eprint={2107.03374},
          archivePrefix={arXiv},
          primaryClass={cs.LG},
          url={https://arxiv.org/abs/2107.03374}, 
    }

@misc{austin2021programsynthesislargelanguage,
          title={Program Synthesis with Large Language Models}, 
          author={Jacob Austin and Augustus Odena and Maxwell Nye and Maarten Bosma and Henryk Michalewski and David Dohan and Ellen Jiang and Carrie Cai and Michael Terry and Quoc Le and Charles Sutton},
          year={2021},
          eprint={2108.07732},
          archivePrefix={arXiv},
          primaryClass={cs.PL},
          url={https://arxiv.org/abs/2108.07732}, 
    }

@misc{kaplan2020scalinglawsneurallanguage,
              title={Scaling Laws for Neural Language Models}, 
              author={Jared Kaplan and Sam McCandlish and Tom Henighan and Tom B. Brown and Benjamin Chess and Rewon Child and Scott Gray and Alec Radford and Jeffrey Wu and Dario Amodei},
              year={2020},
              eprint={2001.08361},
              archivePrefix={arXiv},
              primaryClass={cs.LG},
              url={https://arxiv.org/abs/2001.08361}, 
        }

@inproceedings{xia2024unlockingefficiencylargelanguage,
  title={Unlocking Efficiency in Large Language Model Inference: A Comprehensive Survey of Speculative Decoding},
  author={Xia, Heming and Yang, Zhe and Dong, Qingxiu and Wang, Peiyi and Li, Yongqi and Ge, Tao and Liu, Tianyu and Li, Wenjie and Sui, Zhifang},
  booktitle={ACL (Findings)},
  year={2024},
  doi={10.18653/v1/2024.findings-acl.456}
}

@inproceedings{jain2024livecodebench,
  title={LiveCodeBench: Holistic and Contamination Free Evaluation of Large Language Models for Code},
  author={Jain, Naman and Han, King and Gu, Alex and Li, Wen-Ding and Yan, Fanjia and Zhang, Tianjun and Wang, Sida and Solar-Lezama, Armando and Sen, Koushik and Stoica, Ion},
  booktitle={The Thirteenth International Conference on Learning Representations}
}

@inproceedings{jimenez2024swebench,
  title={SWE-bench: Can Language Models Resolve Real-world Github Issues?},
  author={Jimenez, Carlos E and Yang, John and Wettig, Alexander and Yao, Shunyu and Pei, Kexin and Press, Ofir and Narasimhan, Karthik R},
  booktitle={The Twelfth International Conference on Learning Representations}
}

@misc{aider2024polyglot,
      author = {Aider AI},
      title = {polyglot-benchmark: Coding problems used in aider's polyglot benchmark},
      year = {2024},
      publisher = {GitHub},
      journal = {GitHub repository},
      howpublished = {\url{https://github.com/Aider-AI/polyglot-benchmark}}
    }

@article{kwon2023efficient,
  title={Efficient Memory Management for Large Language Model Serving with PagedAttention},
  author={Kwon, Woosuk and Li, Zhuohan and Zhuang, Siyuan and Sheng, Ying and Zheng, Lianmin and Yu, Cody Hao and Gonzalez, Joseph E and Zhang, Hao and Stoica, Ion},
  journal={CoRR},
  year={2023},
  doi = {10.1145/3600006.3613165},
}

@misc{specforge2025,
      title={SpecForge: Train speculative decoding models effortlessly},
      author={Shenggui Li and Yikai Zhu and Chao Wang and Fan Yin and Shuai Shi and Yubo Wang and Yi Zhang and Yingyi Huang and Haoshuai Zheng and Yineng Zhang},
      year={2025},
      publisher={GitHub},
      howpublished={\url{https://github.com/sgl-project/specforge}},
    }

@article{li2023taco,
  title={TACO: Topics in Algorithmic COde generation dataset},
  author={Li, Rongao and Fu, Jie and Zhang, Bo-Wen and Huang, Tao and Sun, Zhihong and Lyu, Chen and Liu, Guang and Jin, Zhi and Li, Ge},
  journal={CoRR},
  year={2023}
}

@inproceedings{pan2025training,
  title={Training Software Engineering Agents and Verifiers with SWE-Gym},
  author={Pan, Jiayi and Wang, Xingyao and Neubig, Graham and Jaitly, Navdeep and Ji, Heng and Suhr, Alane and Zhang, Yizhe},
  booktitle={Forty-second International Conference on Machine Learning}
}

@inproceedings{elhoushi2024layerskip,
  title={Layerskip: Enabling early exit inference and self-speculative decoding},
  author={Elhoushi, Mostafa and Shrivastava, Akshat and Liskovich, Diana and Hosmer, Basil and Wasti, Bram and Lai, Liangzhen and Mahmoud, Anas and Acun, Bilge and Agarwal, Saurabh and Roman, Ahmed and others},
  booktitle={Proceedings of the 62nd Annual Meeting of the Association for Computational Linguistics (Volume 1: Long Papers)},
  pages={12622--12642},
  year={2024}
}

@inproceedings{miao2024specinfer,
  title={Specinfer: Accelerating large language model serving with tree-based speculative inference and verification},
  author={Miao, Xupeng and Oliaro, Gabriele and Zhang, Zhihao and Cheng, Xinhao and Wang, Zeyu and Zhang, Zhengxin and Wong, Rae Ying Yee and Zhu, Alan and Yang, Lijie and Shi, Xiaoxiang and others},
  booktitle={Proceedings of the 29th ACM International Conference on Architectural Support for Programming Languages and Operating Systems, Volume 3},
  pages={932--949},
  doi = {10.1145/3620666.3651335},
  year={2024}
}

@inproceedings{yang2025hass,
  title={Learning Harmonized Representations for Speculative Sampling},
  author={Zhang, Lefan and Wang, Xiaodan and Huang, Yanhua and Xu, Ruiwen},
  booktitle={The Thirteenth International Conference on Learning Representations}
}

@inproceedings{santilli2023accelerating,
   title={Accelerating Transformer Inference for Translation via Parallel Decoding},
   url={http://dx.doi.org/10.18653/v1/2023.acl-long.689},
   DOI={10.18653/v1/2023.acl-long.689},
   booktitle={Proceedings of the 61st Annual Meeting of the Association for Computational Linguistics (Volume 1: Long Papers)},
   publisher={Association for Computational Linguistics},
   author={Santilli, Andrea and Severino, Silvio and Postolache, Emilian and Maiorca, Valentino and Mancusi, Michele and Marin, Riccardo and Rodola, Emanuele},
   year={2023} }

@inproceedings{liu2023repobenchbenchmarkingrepositorylevelcode,
  title={RepoBench: Benchmarking Repository-Level Code Auto-Completion Systems},
  author={Liu, Tianyang and Xu, Canwen and McAuley, Julian},
  booktitle={The Twelfth International Conference on Learning Representations}
}

@article{zheng2023judgingllmasajudgemtbenchchatbot,
  title={Judging llm-as-a-judge with mt-bench and chatbot arena},
  author={Zheng, Lianmin and Chiang, Wei-Lin and Sheng, Ying and Zhuang, Siyuan and Wu, Zhanghao and Zhuang, Yonghao and Lin, Zi and Li, Zhuohan and Li, Dacheng and Xing, Eric and others},
  journal={Advances in neural information processing systems},
  volume={36},
  pages={46595--46623},
  year={2023}
}

@misc{cobbe2021trainingverifierssolvemath,
      title={Training Verifiers to Solve Math Word Problems}, 
      author={Karl Cobbe and Vineet Kosaraju and Mohammad Bavarian and Mark Chen and Heewoo Jun and Lukasz Kaiser and Matthias Plappert and Jerry Tworek and Jacob Hilton and Reiichiro Nakano and Christopher Hesse and John Schulman},
      year={2021},
      eprint={2110.14168},
      archivePrefix={arXiv},
      primaryClass={cs.LG},
      url={https://arxiv.org/abs/2110.14168}, 
}

@misc{alpaca,
  author = {Rohan Taori and Ishaan Gulrajani and Tianyi Zhang and Yann Dubois and Xuechen Li and Carlos Guestrin and Percy Liang and Tatsunori B. Hashimoto },
  title = {Stanford Alpaca: An Instruction-following LLaMA model},
  year = {2023},
  publisher = {GitHub},
  journal = {GitHub repository},
  howpublished = {\url{https://github.com/tatsu-lab/stanford_alpaca}},
}

@article{hermann2015teachingmachinesreadcomprehend,
  title={Teaching machines to read and comprehend},
  author={Hermann, Karl Moritz and Kocisky, Tomas and Grefenstette, Edward and Espeholt, Lasse and Kay, Will and Suleyman, Mustafa and Blunsom, Phil},
  journal={Advances in neural information processing systems},
  volume={28},
  year={2015}
}

@article{hu2025assessing,
  title={Assessing and advancing benchmarks for evaluating large language models in software engineering tasks},
  author={Hu, Xing and Niu, Feifei and Chen, Junkai and Zhou, Xin and Zhang, Junwei and He, Junda and Xia, Xin and Lo, David},
  journal={ACM Transactions on Software Engineering and Methodology},
  year={2025},
  publisher={ACM New York, NY},
  doi={10.1145/3786771}
}

@inproceedings{just2014defects4j,
  title={Defects4J: A database of existing faults to enable controlled testing studies for Java programs},
  author={Just, Ren{\'e} and Jalali, Darioush and Ernst, Michael D},
  booktitle={Proceedings of the 2014 international symposium on software testing and analysis},
  pages={437--440},
  year={2014}
}

@inproceedings{cassanocan,
  title={Can It Edit? Evaluating the Ability of Large Language Models to Follow Code Editing Instructions},
  author={Cassano, Federico and Li, Luisa and Sethi, Akul and Shinn, Noah and Brennan-Jones, Abby and Ginesin, Jacob and Berman, Edward and Chakhnashvili, George and Lozhkov, Anton and Anderson, Carolyn Jane and others},
  booktitle={First Conference on Language Modeling},
  year={2023}
}

@inproceedings{ding2023crosscodeevaldiversemultilingualbenchmark,
 author = {Ding, Yangruibo and Wang, Zijian and Ahmad, Wasi and Ding, Hantian and Tan, Ming and Jain, Nihal and Ramanathan, Murali Krishna and Nallapati, Ramesh and Bhatia, Parminder and Roth, Dan and Xiang, Bing},
 booktitle = {Advances in Neural Information Processing Systems},
 doi = {10.52202/075280-2023},
 editor = {A. Oh and T. Naumann and A. Globerson and K. Saenko and M. Hardt and S. Levine},
 pages = {46701--46723},
 publisher = {Curran Associates, Inc.},
 title = {CrossCodeEval: A Diverse and Multilingual Benchmark for Cross-File Code Completion},
 url = {https://proceedings.neurips.cc/paper_files/paper/2023/file/920f2dced7d32ab2ba2f1970bc306af6-Paper-Datasets_and_Benchmarks.pdf},
 volume = {36},
 year = {2023}
}

@inproceedings{muennighoff2024octopackinstructiontuningcode,
  title={Octopack: Instruction tuning code large language models},
  author={Muennighoff, Niklas and Liu, Qian and Zebaze, Armel and Zheng, Qinkai and Hui, Binyuan and Zhuo, Terry Yue and Singh, Swayam and Tang, Xiangru and Von Werra, Leandro and Longpre, Shayne},
  booktitle={NeurIPS 2023 workshop on instruction tuning and instruction following}
}

@misc{nvidia2024tensorrtllm,
  author = {{NVIDIA}},
  title = {TensorRT-LLM: A TensorRT Toolbox for Large Language Model Inference},
  url = {https://github.com/NVIDIA/TensorRT-LLM},
  howpublished={\url{https://github.com/NVIDIA/TensorRT-LLM}},
  year = {2024},
  note = {High-performance inference library}
}

@inproceedings{yang2025longspeclongcontextlosslessspeculative,
  title={LongSpec: Long-Context Lossless Speculative Decoding with Efficient Drafting and Verification},
  author={Yang, Penghui and Du, Cunxiao and Zhang, Fengzhuo and Wang, Haonan and Pang, Tianyu and Du, Chao and An, Bo},
  booktitle={ES-FoMo III: 3rd Workshop on Efficient Systems for Foundation Models},
  year={2025},
  doi={10.18653/v1/2026.acl-long.83}
}

@article{deng2025swe,
  title={SWE-Bench Pro: Can AI Agents Solve Long-Horizon Software Engineering Tasks?},
  author={Deng, Xiang and Da, Jeff and Pan, Edwin and He, Yannis Yiming and Ide, Charles and Garg, Kanak and Lauffer, Niklas and Park, Andrew and Pasari, Nitin and Rane, Chetan and others},
  journal={arXiv preprint arXiv:2509.16941},
  year={2025}
}

@article{guo2025deepseek,
  title={DeepSeek-R1 incentivizes reasoning in LLMs through reinforcement learning},
  author={Guo, Daya and Yang, Dejian and Zhang, Haowei and Song, Junxiao and Wang, Peiyi and Zhu, Qihao and Xu, Runxin and Zhang, Ruoyu and Ma, Shirong and Bi, Xiao and others},
  journal={Nature},
  volume={645},
  number={8081},
  pages={633--638},
  year={2025},
  publisher={Nature Publishing Group UK London},
  doi={10.1038/s41586-025-09422-z}
}

@article{jaech2024openai,
  title={Openai o1 system card},
  author={Jaech, Aaron and Kalai, Adam and Lerer, Adam and Richardson, Adam and El-Kishky, Ahmed and Low, Aiden and Helyar, Alec and Madry, Aleksander and Beutel, Alex and Carney, Alex and others},
  journal={arXiv preprint arXiv:2412.16720},
  year={2024}
}

@article{zhaofastcoder,
  title={FASTCODER: Accelerating Repository-level Code Generation via Efficient Retrieval and Verification},
  author={Zhao, Qianhui and Zhang, Li and Liu, Fang and Lian, Xiaoli and Meng, Qiaoyuanhe and Jiao, Ziqian and Zhou, Zetong and Li, Jia and Shi, Lin}
}

@article{peidingefficientedit,
  title={EFFICIENTEDIT: Accelerating Code Editing via Edit-Oriented Speculative Decoding},
  author={Peiding Wang, Li Zhang and Liu, Fang and Zhu, Yinghao and Xu, Wang and Shi, Lin and Lian, Xiaoli and Li, Minxiao and Shen, Bo and Fu, An}
}

@misc{leetcode,
  title        = {LeetCode},
  author       = {{LeetCode Inc.}},
  howpublished = {\url{https://leetcode.com}},
  year         = {2026},
  note         = {Accessed: 2026-01-27}
}

@misc{vllm_v0120_release,
  author       = {{vLLM Team}},
  title        = {{vLLM} v0.12.0 Release Notes},
  howpublished = {\url{https://github.com/vllm-project/vllm/releases/tag/v0.12.0}},
  year         = {2025},
  month        = dec, 
  note         = {Accessed: 2026-01-27},
  publisher    = {GitHub}
}

@misc{khoshnoodi2024comprehensivesurveyacceleratedgeneration,
      title={A Comprehensive Survey of Accelerated Generation Techniques in Large Language Models}, 
      author={Mahsa Khoshnoodi and Vinija Jain and Mingye Gao and Malavika Srikanth and Aman Chadha},
      year={2024},
      eprint={2405.13019},
      archivePrefix={arXiv},
      primaryClass={cs.CL},
      url={https://arxiv.org/abs/2405.13019}, 
}

@article{liang2024can,
  title={Can gpt-4 replicate empirical software engineering research?},
  author={Liang, Jenny T and Badea, Carmen and Bird, Christian and DeLine, Robert and Ford, Denae and Forsgren, Nicole and Zimmermann, Thomas},
  journal={Proceedings of the ACM on Software Engineering},
  volume={1},
  number={FSE},
  pages={1330--1353},
  year={2024},
  publisher={ACM New York, NY, USA},
  doi = {10.1145/3660767},
}

@inproceedings{huang2023empirical,
  title={An empirical study on fine-tuning large language models of code for automated program repair},
  author={Huang, Kai and Meng, Xiangxin and Zhang, Jian and Liu, Yang and Wang, Wenjie and Li, Shuhao and Zhang, Yuqing},
  booktitle={2023 38th IEEE/ACM International Conference on Automated Software Engineering (ASE)},
  pages={1162--1174},
  year={2023},
  organization={IEEE},
  doi={10.1109/ASE56229.2023.00181}
}

@article{sun2025empirical,
  title={Empirical Evaluation of Large Language Models in Automated Program Repair},
  author={Sun, Jiajun and Li, Fengjie and Qi, Xinzhu and Zhang, Hongyu and Jiang, Jiajun},
  journal={arXiv preprint arXiv:2506.13186},
  year={2025}
}

@article{shang2025large,
  title={A large-scale empirical study on fine-tuning large language models for unit testing},
  author={Shang, Ye and Zhang, Quanjun and Fang, Chunrong and Gu, Siqi and Zhou, Jianyi and Chen, Zhenyu},
  journal={Proceedings of the ACM on Software Engineering},
  volume={2},
  number={ISSTA},
  doi = {10.1145/3728951},
  pages={1678--1700},
  year={2025},
  publisher={ACM New York, NY, USA}
}

@inproceedings{jiang2023impact,
  title={Impact of code language models on automated program repair},
  author={Jiang, Nan and Liu, Kevin and Lutellier, Thibaud and Tan, Lin},
  booktitle={2023 IEEE/ACM 45th International Conference on Software Engineering (ICSE)},
  pages={1430--1442},
  year={2023},
  organization={IEEE},
  doi = {10.1109/ICSE48619.2023.00125},

}

@inproceedings{zhou2024out,
  title={Out of sight, out of mind: Better automatic vulnerability repair by broadening input ranges and sources},
  author={Zhou, Xin and Kim, Kisub and Xu, Bowen and Han, DongGyun and Lo, David},
  booktitle={Proceedings of the IEEE/ACM 46th international conference on software engineering},
  pages={1--13},
  year={2024},
  doi = {10.1145/3597503.3639222},
}

@article{wang2025empirical,
  title={An Empirical Study of Knowledge Distillation for Code Understanding Tasks},
  author={Wang, Ruiqi and Yang, Zezhou and Gao, Cuiyun and Xia, Xin and Liao, Qing},
  journal={arXiv preprint arXiv:2508.15423},
  year={2025}
}

@INPROCEEDINGS{afrin2025quantization,
  author={Afrin, Saima and Xu, Bowen and Mastropaolo, Antonio},
  booktitle={2025 IEEE International Conference on Software Maintenance and Evolution (ICSME)}, 
  title={Is Quantization a Deal-Breaker? Empirical Insights From Large Code Models}, 
  year={2025},
  volume={},
  number={},
  pages={1-13},
  doi={10.1109/ICSME64153.2025.00049}}

@article{melin2024precision,
  title={Precision or peril: Evaluating code quality from quantized large language models},
  author={Melin, Eric L and Torek, Adam J and Eisty, Nasir U and Kennington, Casey},
  journal={arXiv preprint arXiv:2411.10656},
  year={2024},
  doi = {10.1145/3786175.3788343},
}

@article{haque2025quantization,
  title={How Quantization Impacts Privacy Risk on LLMs for Code?},
  author={Haque, Md Nazmul and Yang, Hua and Yang, Zhou and Xu, Bowen},
  journal={arXiv preprint arXiv:2508.00128},
  year={2025}
}

@article{siam2026exploring,
  title={Exploring the Limits of Pruning: Task-Specific Neurons, Model Collapse, and Recovery in Task-Specific Large Language Models},
  author={Siam, MK and Tausif-Ul-Islam, Md and Khan, Md Reshad Romim and Hossain, Mohammed Ali and Amin, Mushfiqul and Khan, Labib Hasan and Farhan, Niloy and Sadeque, Farig},
  journal={arXiv preprint arXiv:2604.27115},
  year={2026}
}

@article{xu2024aligning,
  title={Aligning the Objective of LLM-based Program Repair},
  author={Xu, Junjielong and Fu, Ying and Tan, Shin Hwei and He, Pinjia},
  journal={arXiv preprint arXiv:2404.08877},
  year={2024},
  doi = {10.1109/ICSE55347.2025.00169},
}

@inproceedings{chen2024code,
  title={Code search is all you need? improving code suggestions with code search},
  author={Chen, Junkai and Hu, Xing and Li, Zhenhao and Gao, Cuiyun and Xia, Xin and Lo, David},
  booktitle={Proceedings of the IEEE/ACM 46th international conference on software engineering},
  pages={1--13},
  year={2024}
}

@inproceedings{chen2025reasoning,
  title={Reasoning Runtime Behavior of a Program with LLM: How Far are We?},
  author={Chen, Junkai and Pan, Zhiyuan and Hu, Xing and Li, Zhenhao and Li, Ge and Xia, Xin},
  booktitle={2025 IEEE/ACM 47th International Conference on Software Engineering (ICSE)},
  pages={1869--1881},
  year={2025},
  organization={IEEE}
}

@inproceedings{chen2026securevibebench,
  title={Securevibebench: Benchmarking secure vibe coding of ai agents via reconstructing vulnerability-introducing scenarios},
  author={Chen, Junkai and Huang, Huihui and Lyu, Yunbo and An, Junwen and Shi, Jieke and Yang, Chengran and Zhang, Ting and Tian, Haoye and Li, Yikun and Li, Zhenhao and others},
  booktitle={Proceedings of the 64th Annual Meeting of the Association for Computational Linguistics (Volume 1: Long Papers)},
  pages={24144--24168},
  year={2026}
}

@inproceedings{yang2026securepair,
  title={SeCuRepair: Semantics-Aligned, Curriculum-Driven, and Reasoning-Enhanced Vulnerability Repair Framework},
  author={Yang, Chengran and Zhang, Ting and Jiang, Jinfeng and Zhou, Xin and Tian, Haoye and Du, Mingzhe and Shi, Jieke and Chen, Junkai and Li, Yikun and Ouh, Eng Lieh and others},
  booktitle={Proceedings of the 64th Annual Meeting of the Association for Computational Linguistics (Volume 1: Long Papers)},
  pages={32108--32123},
  year={2026}
}

\end{document}